\documentclass[preprint2]{proto}
\usepackage{times}
\newcommand{\refs}{\par\noindent\hangindent=1pc\hangafter=1}
\def\ltsima{$\; \buildrel < \over \sim \;$}
\def\simlt{\lower.5ex\hbox{\ltsima}}
\def\gtsima{$\; \buildrel > \over \sim \;$}
\def\simgt{\lower.5ex\hbox{\gtsima}}

\def\arcsec{$^{\prime\prime}$}

\begin{document}

\title{\textbf{\LARGE From Filamentary Networks to Dense Cores in Molecular Clouds:
Toward a New Paradigm for Star Formation}}

\author {\textbf{\large Philippe Andr\'e}}
\affil{\small\em Laboratoire d'Astrophysique de Paris-Saclay}

\author {\textbf{\large James Di Francesco}}
\affil{\small\em National Research Council of Canada}

\author {\textbf{\large Derek Ward-Thompson}}
\affil{\small\em University of Central Lancashire}

\author {\textbf{\large Shu-ichiro Inutsuka}}
\affil{\small\em Nagoya University}

\author {\textbf{\large Ralph E. Pudritz}}
\affil{\small\em McMaster University}

\author {\textbf{\large Jaime Pineda}}
\affil{\small\em University of Manchester and European Southern Observatory}

\affil{\em }


\begin{abstract}
\baselineskip = 11pt
\leftskip = 0.65in 
\rightskip = 0.65in
\parindent=1pc
{\small 
Recent studies of the nearest star-forming clouds of the Galaxy at submillimeter wavelengths with the 
$Herschel$ Space Observatory have provided us with unprecedented images of the initial and boundary 
conditions of the star formation process. The $Herschel$ results emphasize the role of interstellar filaments 
in the star formation process and connect remarkably well with nearly a decade's worth of numerical simulations 
and theory that have consistently shown that the ISM should be highly filamentary on all scales and 
star formation is intimately related to self-gravitating filaments.
In this review, we trace how the apparent complexity of cloud structure and star formation is governed 
by relatively simple universal processes - from filamentary clumps to galactic scales.
We emphasize two crucial and complementary aspects:  (i) the key observational results obtained with 
$Herschel$ over the past three years, along with relevant new results obtained from the ground on the 
kinematics of interstellar structures, and (ii) the key existing theoretical models and the many numerical 
simulations of interstellar cloud structure and star formation. 
We then synthesize a comprehensive physical picture that arises from the confrontation of these 
observations and simulations.  
 \\~\\~\\~}

\end{abstract}  

\section{\textbf{INTRODUCTION}}

The physics controlling the earliest  
phases of star formation is not yet well understood. Improving our global understanding of these phases is crucial for gaining insight into the general inefficiency of the star formation process, the global rate of star formation on galactic scales, the origin of stellar masses, and  
the birth of planetary systems. 

\begin{figure*}
 \epsscale{2}
\plotone{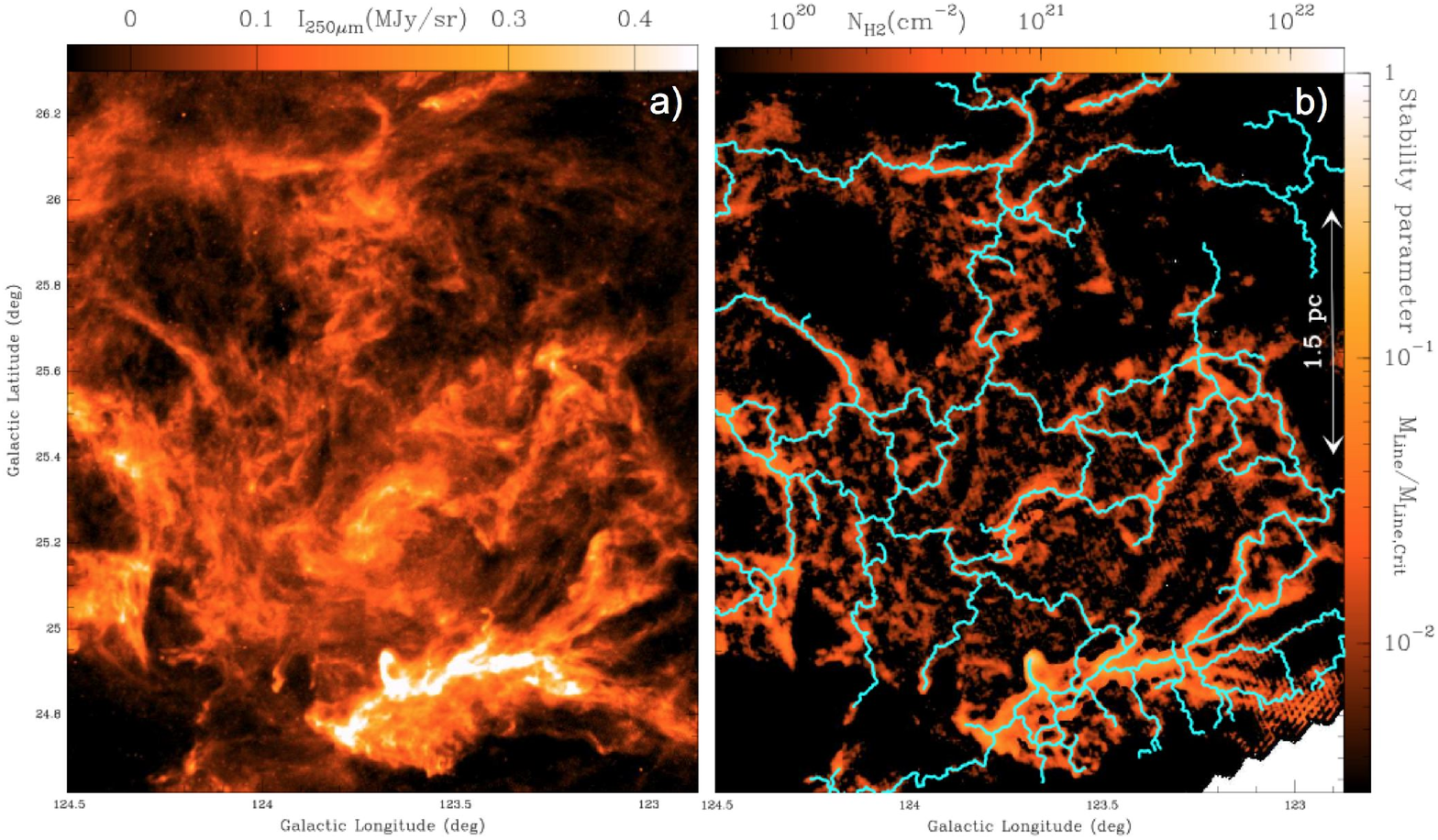}
 \caption{\small 
 {\bf(a)}  $Herschel$/SPIRE 250~$\mu$m dust continuum map of a portion of the Polaris flare translucent cloud 
(e.g., {\em Miville-Desch\^enes et al.,} 2010, {\em Ward-Thompson et al.,} 2010).
{\bf(b)}  Corresponding column density map derived from  $Herschel$ data (e.g., {\em Andr\'e et al.,} 2010). 
The contrast of the filaments has been enhanced using a curvelet transform (cf. {\em Starck et al.,} 2003). 
The skeleton of the filament network identified 
with the DisPerSE algorithm ({\em Sousbie,} 2011) is shown in light blue. 
A similar pattern is found with other algorithms such as {\it getfilaments} ({\em Men'shchikov,} 2013). 
Given the typical width $\sim $~0.1~pc of the filaments  ({\em Arzoumanian et al.,} 2011 -- see Fig.~\ref{histo_width} below), this column density map 
is equivalent to a {\it map of the mass per unit length along the filaments} (see color scale on the right). 
 }
 \label{Polaris}     

\vspace{0.75cm}
\epsscale{1.9}
\plotone{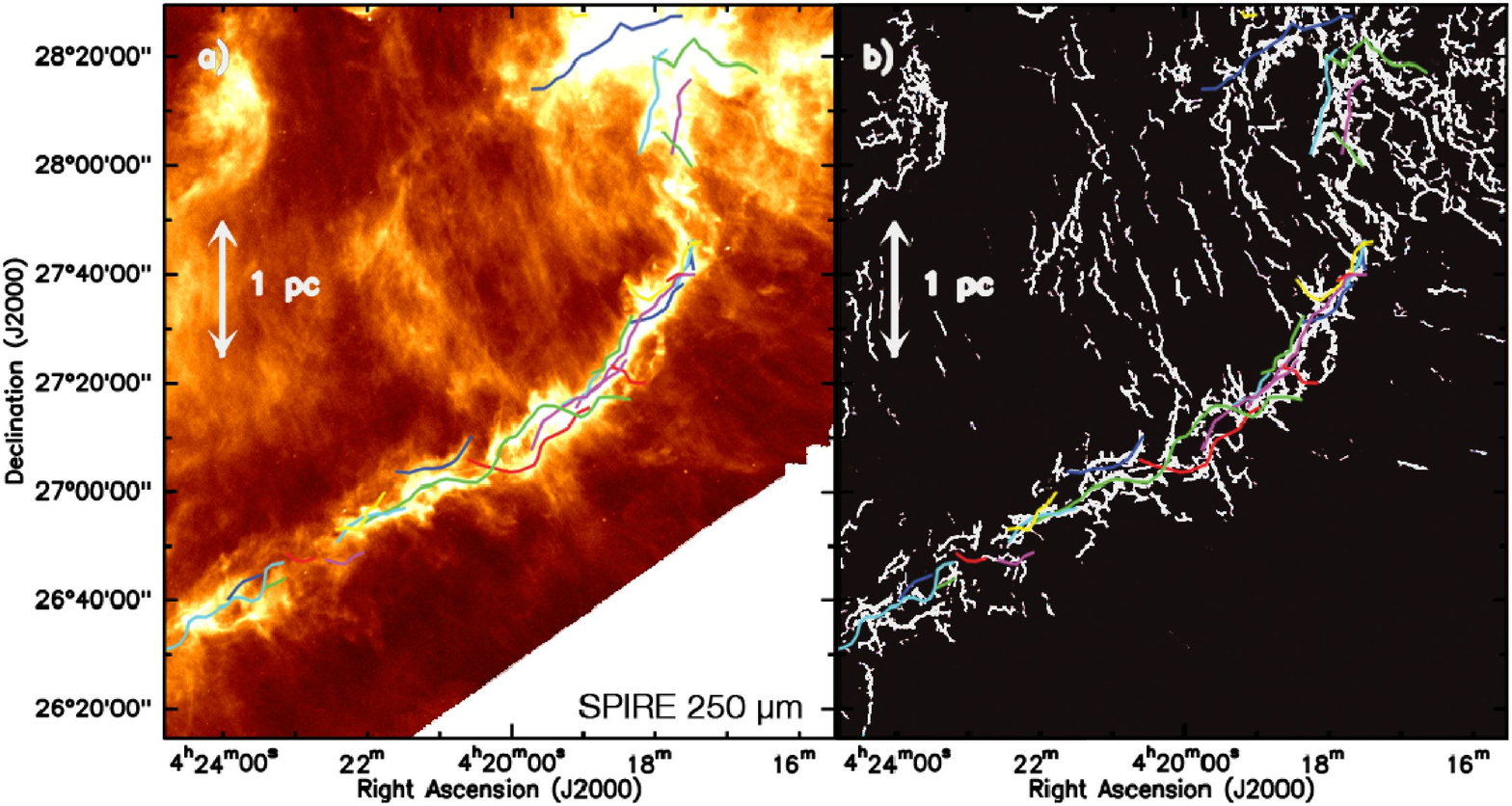}
 \caption{\small 
 {\bf(a)} $Herschel$/SPIRE 250 $\mu$m dust continuum image of the B211/B213/L1495 region in Taurus ({\em Palmeirim et al.,} 2013). 
The colored curves display the velocity-coherent ``fibers'' identified within the B213/B211 filament 
by {\em Hacar et al.} (2013) using C$^{18}$O(1--0) observations. 
    {\bf(b)} Fine structure of the $Herschel$/SPIRE 250 $\mu$m dust continuum emission from the B211/B213 filament 
obtained by applying the multi-scale algorithm {\it getfilaments} ({\em Men'shchikov,} 2013) to the 250 $\mu$m image shown in panel (a). 
Note the faint striations perpendicular to the main filament and 
the excellent correspondence between the small-scale structure of the dust continuum filament and the bundle 
of velocity-coherent fibers traced by {\em Hacar et al.} (2013) in C$^{18}$O (same colored curves as in (a)).
}  
\label{B211_fil_fibers}
\end{figure*}

Since PPV seven years ago, one area that has seen the most dramatic advances has been the characterization of the link between star formation and the structure of the cold interstellar medium (ISM). In particular, extensive studies of the nearest star-forming clouds of our Galaxy with the $Herschel$ Space Observatory have provided us with unprecedented images of the initial and boundary conditions of the star formation process (e.g., Fig.~\ref{Polaris}). 
The $Herschel$ images reveal an intricate network of filamentary structures in every interstellar cloud. The observed filaments share common properties, such as their central widths,  
but only the densest filaments contain prestellar cores, the seeds of future stars. Overall, the $Herschel$ data, as well as other observations from, e.g., near-IR extinction studies, favor a scenario in which interstellar filaments and prestellar cores represent two key steps in the star formation process. 
First large-scale supersonic flows compress the gas,  
giving rise to a universal web-like filamentary structure in the ISM. 
Next, gravity takes over and controls the further fragmentation of filaments into prestellar cores and ultimately protostars.

The new observational results connect remarkably well with nearly a decade's worth of numerical simulations and theory that have consistently shown the ISM should be highly filamentary on all scales and star formation is intimately connected with self-gravitating filaments (e.g., Fig.~\ref{Simulations1} in \S~5 below). 
The observations set strong constraints on  
models for the growth of structure in the ISM leading to the formation of young stellar populations and star clusters. Numerical simulations now successfully include turbulence, gravity, a variety of cooling processes, MHD, and most recently, radiation and radiative feedback from massive stars. These numerical advances have been essential in testing and developing new insights into the physics of filaments and star formation, 
including the formation, fragmentation, and further evolution of filaments through accretion, and the central role of filaments 
in the rapid gathering of gas into cluster forming, dense regions.  

\bigskip

\centerline{\textbf{ 2. UNIVERSALITY OF THE FILAMENTARY }}
\centerline{\textbf{ STRUCTURE OF THE COLD ISM}}
\bigskip

\noindent
\textbf{ 2.1 Evidence of interstellar filaments prior to $Herschel$ }
\smallskip

The presence of parsec-scale filamentary structures in nearby interstellar clouds and their potential 
importance for star formation have been pointed out by many authors for more than three decades. 
For example, {\em Schneider and Elmegreen} (1979) discussed the properties of 23 elongated dark nebulae, 
visible on optical plates and showing evidence of marked condensations or internal globules, 
which they named ``globular filaments.''  
High-angular resolution observations of HI toward the Riegel-Crutcher cloud by {\em McClure-Griffiths et al.} 
(2006) also revealed an impressive network of aligned HI filaments.  In this case, the filaments were
observed in HI self absorption (HISA) but with a column density ($A_V$ $<$ 0.5 mag) that is not enough
to shield CO from photodissociation.   
These HI filaments appear aligned with the ambient magnetic field, 
suggesting they are magnetically dominated.
Tenuous CO 
filaments were also observed in diffuse molecular gas by {\em Falgarone et al.} (2001) 
and {\em Hily-Blant and Falgarone} (2007).

Within star-forming molecular gas, two nearby complexes were noted to have prominent filamentary 
structure in both CO and dust maps: the Orion A cloud (e.g., {\em Bally et al.,} 1987; {\em Chini et al.,} 1997; {\em Johnstone 
and Bally,}  1999) and the Taurus cloud (e.g., {\em Abergel et al.,} 1994; {\em Mizuno et al.,} 1995; {\em Hartmann,} 2002; {\em Nutter et al.,} 2008; {\em Goldsmith 
et al.,} 2008).  Other well-known examples include the 
molecular clouds in Musca-Chamaeleon (e.g., {\em Cambr\'esy,} 1999), 
Perseus (e.g., {\em Hatchell et al.,} 2005), and S106 (e.g., {\em Balsara et al.,} 2001). 
More distant ``infrared dark clouds'' (IRDCs) identified at 
mid-infrared wavelengths with $ISO$, $MSX$, and $Spitzer$ (e.g., {\em P\'erault et al.,} 1996; {\em Egan et al.,} 1998; 
{\em Peretto and Fuller,} 2009; {\em Miettinen and Harju,} 2010), some of which are believed to be the birthplaces of massive 
stars (see chapter by {\em Tan et al.}), also have clear filamentary morphologies.  Collecting and comparing observations available prior 
to $Herschel$, {\em Myers} (2009) noticed that young stellar groups and clusters are frequently 
associated with dense ``hubs'' radiating multiple filaments made of lower column density material. 

For the purpose of this review, we will define a ``filament'' as any elongated ISM structure with an aspect 
ratio larger than $\sim \, $5--10 that is significantly overdense with respect to its surroundings. 
The results of Galactic imaging surveys carried out with the $Herschel$ Space Observatory ({\em Pilbratt et al.,} 2010) 
between late 2009 and early 2013 now demonstrate that such filaments are truly ubiquitous in the cold ISM 
(e.g,. \S ~2.2 below), present a high degree of universality in their properties  (e.g., \S ~2.5), 
and likely play a key role in the star formation process (see \S ~6). 

\bigskip
\noindent
\textbf{ 2.2 Ubiquity of filaments in $Herschel$ imaging surveys }
\medskip
 
{\it Herschel}\/ images 
provide key information on the structure of molecular clouds over
spatial scales ranging from the sizes of entire cloud complexes ($\geq$ 10 
pc) down to the sizes of individual dense cores ($<$ 0.1 pc).  While many
interstellar clouds were already known to exhibit large-scale filamentary
structures long before (cf. \S ~2.1 above), one of the most spectacular early 
findings from {\it Herschel}\/ continuum observations was that the cold ISM 
(e.g., individual molecular clouds, GMCs, and the Galactic Plane) is highly
structured with filaments pervading clouds (e.g., see {\em Andr\'e et al.,} 2010; 
{\em MenÕshchikov et al.,} 2010; {\em Molinari et al.,} 2010; {\em Henning et al.,} 2010; {\em Motte 
et al.,} 2010).  These ubiquitous structures were seen by {\it Herschel}\/ for 
the first time due to its extraordinary sensitivity to thermal dust emission 
both at high resolution and over (larger) scales previously inaccessible 
from the ground.  

Filamentary structure is  
omnipresent in every cloud 
observed with {\it Herschel}, irrespective of its star-forming content (see also chapter by {\em Molinari et al.}).  
For example, Fig.~\ref{Polaris}a shows the 250 $\mu$m continuum emission map of the Polaris 
Flare, a translucent, non-star-forming cloud ({\em Ward-Thompson et al.,} 2010; 
{\em Miville-Desch\^enes et al.,} 2010).  
Figure~\ref{Polaris}b shows 
a {\it Herschel}-derived column density map of the same cloud, 
appropriately 
filtered to emphasize the filamentary structure in the data. 
In both panels, filaments are clearly 
seen across the entire cloud, though no star formation has occurred.  This 
omnipresent structure 
suggests the formation of filaments precedes star 
formation in the cold ISM, and is tied to processes acting within the 
clouds themselves.

\begin{figure}[ht]
\epsscale{1}
\plotone{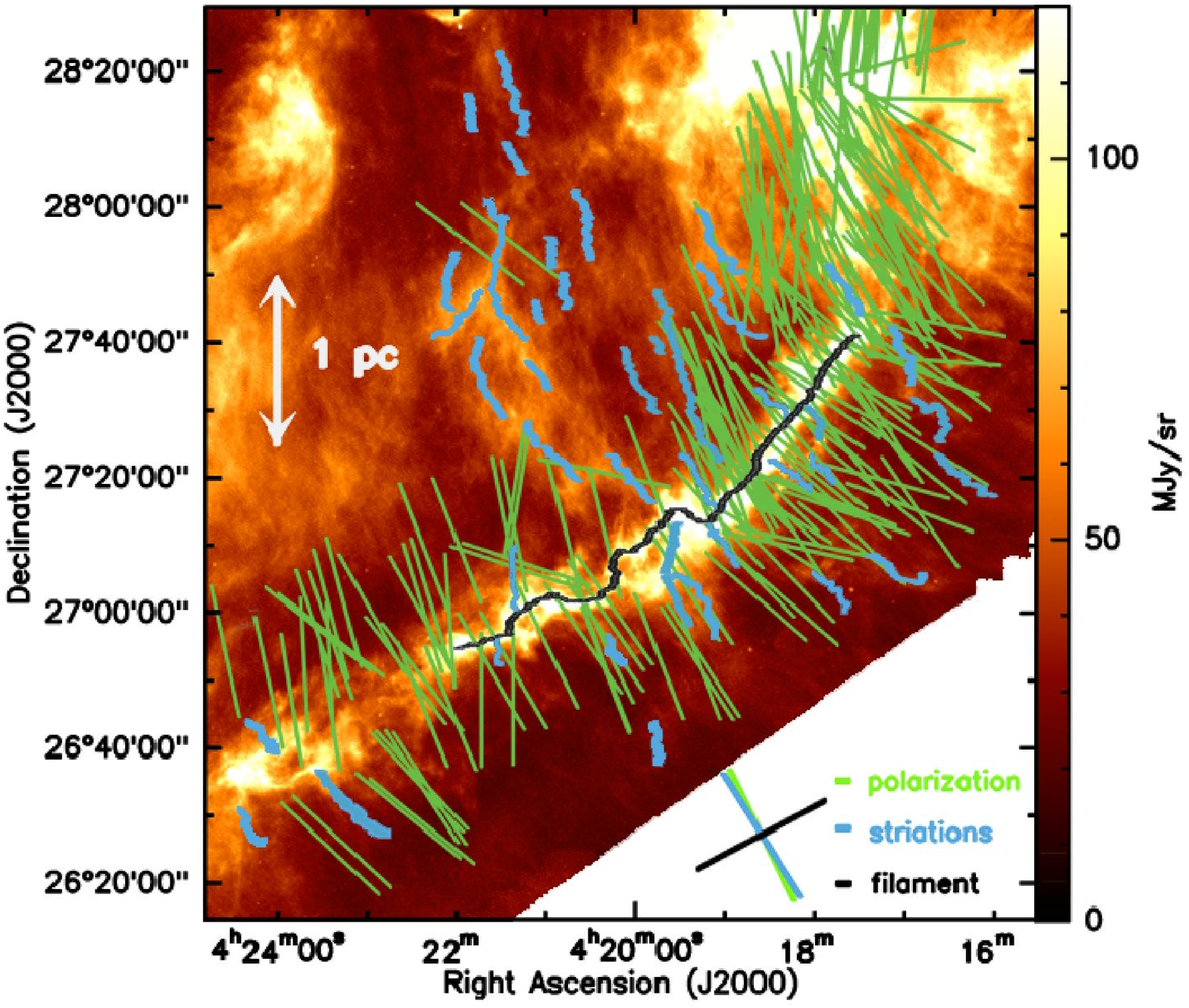} 
 \caption{\small 
 Display of optical and infrared polarization vectors from (e.g., {\em Heyer et al.,} 2008; {\em Chapman et al.,} 2011)
    tracing the magnetic field orientation, overlaid on the $Herschel$/SPIRE 250 $\mu$m image of the 
    B211/B213/L1495 region in Taurus ({\em Palmeirim et al.,} 2013 -- see Fig.~\ref{B211_fil_fibers}a).
   The plane-of-the-sky projection of the magnetic field appears to be oriented perpendicular to the B211/B213 filament and roughly aligned with the 
   general direction of the striations overlaid in blue. 
   A very similar pattern is observed in the Musca cloud ({\em Cox et al.,} in prep.). 
} 
\label{B211_fil}
\end{figure}

\bigskip
\noindent
\textbf{ 2.3 Common patterns in the organization of filaments }
\medskip

The filaments now seen in the cold ISM offer clues about the nature of 
the processes in play within molecular clouds.  First, we note in the clouds 
shown in Fig.~\ref{Polaris} and Fig.~\ref{B211_fil_fibers} (i.e., Polaris and Taurus), 
as well as other clouds, that filaments are typically very long, with lengths of $\sim$1 pc 
or more,   
up to several tens of pc in the case of some IRDCs (e.g., {\em Jackson et al.,} 2010; {\em Beuther et al.,} 2011). 
Despite small-scale deviations, filaments are in 
general quite linear (uni-directional) over their lengths, with typically 
minimal overall curvature and no sharp changes in overall direction.  Moreover, 
many (though not all) filaments appear co-linear in direction to the longer 
extents of their host clouds.  Indeed, some clouds appear globally to be 
filamentary but also contain within themselves distinct populations of 
(sub-)filaments (e.g., IC 5146 or NGC 6334; see {\em Arzoumanian et al.,} 2011; {\em Russeil et 
al.,} 2013).  What is striking about these characteristics is 
that they persist from 
cloud to cloud, though presumably filaments (and their host clouds) are 3-D 
objects with various orientations seen in projection on the sky.  Nevertheless, 
these common traits suggest filaments originate from processes acting over 
the large scales of their host cloud (e.g., large turbulent modes).  
 
What controls the organization of filaments? 
{\em Hill et al.,} (2011) noted how filament networks varied within 
Vela C, itself quite a linear cloud.  They found filaments were arranged 
in more varied directions (i.e., disorganized) within ``nests" in outer, 
lower column density locations but filaments appeared more uni-directional 
within ``ridges" in inner, higher column density locations.  Indeed, {\em Hill 
et al.} found that a greater concentration of mass within filaments is seen 
in ridges vs. nests.  {\em Hill et al.} argued (from comparing column density 
probability functions) that these differences were due to the relative 
influences of turbulence and gravity in various locations within Vela C, 
with the former and latter being dominant in nests and ridges respectively.  
Similar trends can be seen in other clouds.  For example, we note in 
Fig.~\ref{Polaris} the relatively disorganized web-like network of filaments seen in the 
generally lower column density Polaris Flare, whose filaments 
are likely unbound 
and where 
turbulence likely dominates.  On the other hand, we note in Fig.~\ref{B211_fil_fibers} the 
dominant and more unidirectional B211/B213 filament in Taurus which, as judged from 
$Herschel$ data, 
is dense enough to be self-gravitating ({\em Palmeirim et al.,} 2013). 

\begin{figure*}
 \epsscale{2}
 \plotone{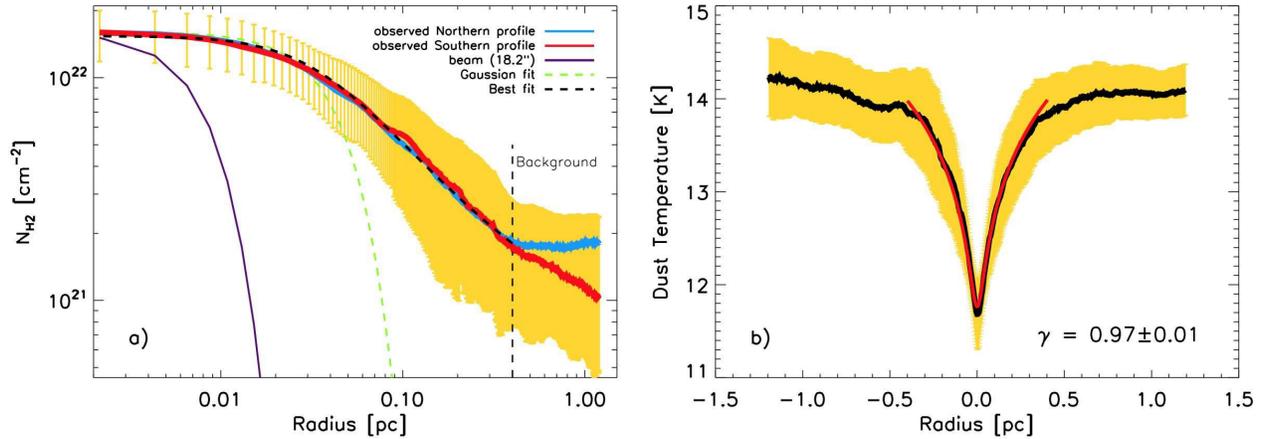}
  \vspace{-0.20cm}
 \caption{\small 
 {\bf(a)}  Mean radial column density profile observed perpendicular to the B213/B211 filament in Taurus ({\em Palmeirim et al.,} 2013), 
for both the Northern (blue curve) and the Southern part (red curve) of the filament.
The yellow area shows the ($\pm 1\sigma$) dispersion of the distribution of radial profiles along the filament.
The inner solid purple curve shows the effective 18\arcsec ~HPBW resolution
(0.012~pc at 140~pc) of  the $Herschel$ column density map used to construct the profile. 
The dashed black curve shows the best-fit Plummer model (convolved with the 18\arcsec ~beam) described by Eq.~(1) 
with $p$=2.0$\pm$0.4 and a diameter $2\times R_{\rm \rm flat} = 0.07 \pm 0.01$~pc, which matches the data very well for $r$$\leq$0.4\,pc, 
{\bf(b)}  Mean dust temperature profile measured perpendicular to the B213/B211 filament in Taurus. 
The solid red curve shows the best polytropic model temperature profile obtained by assuming 
$T_{\rm gas} = T_{\rm dust}$ and 
that the filament has a density profile given 
by the Plummer model shown in the left panel (with $p = 2$) and obeys a polytropic equation of state, $P \propto \rho^{\gamma}$
[and thus $T(r) \propto \rho(r)^{(\gamma-1)}$]. This best fit corresponds to a polytropic index  $\gamma$=0.97$\pm$0.01 
(see {\em Palmeirim et al.,} 2013 for further details).
}
\label{B211_prof} 
\end{figure*}

Although direct observational constraints on the magnetic field inside molecular clouds remain scarce, 
an emerging trend is that dense (self-gravitating) filaments tend to be {\it perpendicular} to the direction of the local magnetic field,
while low-density (unbound) filaments or striations tend to be {\it parallel} to the local magnetic field 
(cf. Fig.~\ref{B211_fil} and {\em Peretto et al.,} 2012; {\em Palmeirim et al.,} 2013; {\em Cox et al.}, in prep. -- see also related chapter by {\em H.-B. Li et al.} 
and caveats discussed by {\em Goodman et al.,} 1990). 

\bigskip
\noindent
\textbf{ 2.4 Density and temperature profiles of filaments}
\bigskip

Detailed analysis of resolved filamentary 
column density profiles 
derived from $Herschel$ data (e.g., {\em Arzoumanian et al.,} 2011; {\em Juvela et al.,} 2012; {\em Palmeirim et al.,} 2013;  and Fig.~\ref{B211_prof}) 
suggests that the shape of filament radial profiles is quasi-universal and 
well described by a Plummer-like function of the form (cf. {\em Whitworth and Ward-Thompson,} 2001; {\em Nutter et al.,} 2008; {\em Arzoumanian et al.,} 2011):
\begin{equation}
 \rho_{p}(r) = \frac{\rho_{c}}{\left[1+\left({r/R_{\rm \rm flat}}\right)^{2}\right]^{p/2}}
\end{equation}

\noindent
for the density profile, equivalent to:

$$ \Sigma_{p}(r) = A_{p}\,  \frac{\rho_{\rm c}R_{\rm \rm flat}}{\left[1+\left({r/R_{\rm \rm flat}}\right)^{2}\right]^{\frac{p-1}{2}}}~~ \ \ (1')$$

\noindent 
for the column density profile, 
where $\rho_{c}$ is the central density of the filament, $R_{\rm \rm flat}$ is the radius of the flat inner region, $p \approx 2$ 
is the power-law exponent at large radii ($r$$>>$$R_{\rm flat}$), and $A_p $ 
is a finite constant factor 
which includes the effect of the filament's inclination angle to the plane of sky.
Note that the density structure of an isothermal gas cylinder in hydrostatic equilibrium follows Eq.~(1) with $p = 4$ ({\em Ostriker,} 1964). 
These recent results with $Herschel$ on the radial density profiles of filaments, illustrated in Fig.~\ref{B211_prof},  
now confirm on a very strong statistical basis 
similar findings obtained on just a handful of filamentary structures from, e.g., near-infrared extinction and ground-based submillimeter 
continuum data (e.g., {\em Lada et al.,} 1999; {\em Nutter et al.,} 2008; {\em Malinen et al.,} 2012).

One possible interpretation for why the power-law exponent of the density profile is $p \approx 2$ at large radii, and 
not $p = 4$ as in the isothermal equilibrium solution, is that dense filaments are not strictly isothermal but better 
described 
by a polytropic equation of state, $P \propto \rho^{\gamma}$ or $T \propto \rho^{\gamma-1}$ with $\gamma \simlt 1$ 
(see {\em Palmeirim et al.,} 2013 and Fig.~\ref{B211_prof}b). 
Models of polytropic cylindrical filaments undergoing gravitational contraction indeed have density profiles scaling 
as $\rho \propto r^{-\frac{2}{2-\gamma}}$ at large radii ({\em Kawachi and Hanawa,} 1998; {\em Nakamura and Umemura,} 1999).
For $\gamma$ values close to unity, the model density profile thus approaches $\rho \propto r^{-2}$, 
in agreement with the $Herschel$ observations. 
However, filaments may be more dynamic systems undergoing accretion and flows of 
various kinds (see \S ~4 and Fig.~\ref{velo_disp} below), 
which we take up in Sect.~5.

\bigskip
\noindent
\textbf{ 2.5 Quasi-universal inner width of interstellar filaments}
\smallskip

\begin{figure}[ht]
 \epsscale{1}
\plotone{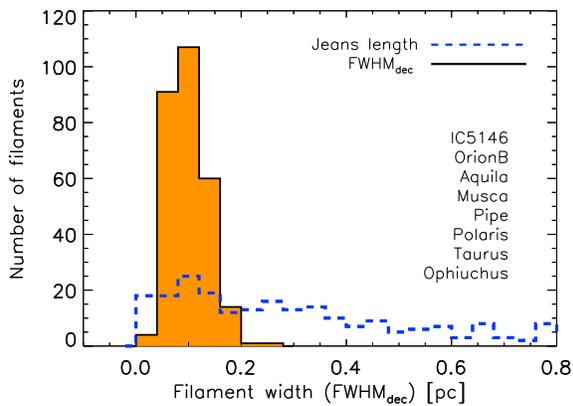}
 \caption{\small 
Histogram of deconvolved FWHM widths for a sample of 278 filaments in 8 nearby regions of the Gould Belt, 
all observed with $Herschel$ (at 
resolutions ranging from $\sim 0.01$~pc to $\sim 0.04$~pc) 
and analyzed in the same way ({\em Arzoumanian et al.}, in prep. --  see {\em Arzoumanian et al.,} 2011 for initial results 
on a subsample of 90 filaments in 3 clouds). 
The distribution of filament widths is narrow 
with a median value of 0.09~pc and a standard deviation of 0.04~pc (equal to the bin size). In contrast, the distribution of Jeans lengths corresponding 
to the central column densities of the filaments (blue dashed histogram) is much broader.  
 }
\label{histo_width} 
\end{figure}

Remarkably, when averaged over the length of the filaments, 
the diameter $2 \times R_{\rm flat}$ of the flat inner plateau in the radial profiles is  
a roughly constant $\sim 0.1$~pc for all filaments, at least in the nearby clouds of Gould's Belt  (cf. {\em Arzoumanian et al.,}  
2011).  For illustration, Fig.~\ref{histo_width} shows that interstellar filaments from eight nearby clouds are 
characterized by a  
narrow distribution of inner FWHM widths centered at about 0.1~pc. 
In contrast, the range of filament column densities probed with $Herschel$ spans 2 to 3 orders 
of magnitude (see Fig.~3 of {\em Arzoumanian et al.,} 2001).
 
The origin of this quasi-universal inner width of interstellar filaments is not yet well understood (see discussion in  \S ~6.5 below). 
A possible hint is that it roughly matches the scale below which the {\em Larson} (1981) 
power-law linewidth vs. size relation breaks down in molecular clouds (cf. {\em Falgarone et al.,} 2009) and a transition to ``coherence'' 
is observed between supersonic and subsonic turbulent gas motions (cf. {\em Goodman et al.,} 1998 and \S ~4.2 below). 
This may suggest that the filament width 
corresponds to the sonic scale below which interstellar turbulence becomes subsonic in diffuse, non-star-forming molecular gas 
(cf. {\em Padoan et al.,} 2001).  
For the densest (self-gravitating) filaments which are expected to radially contract with time (see \S ~5.1 below), this  
interpretation
is unlikely to work, however, and we speculate in \S ~6.5 that {\it accretion} of background cloud material 
(cf. \S ~4) plays a key role.

\bigskip
\noindent
\textbf{ 2.6 Converging filaments and cluster formation}
\bigskip

\begin{figure*}   
\setlength{\unitlength}{1mm}  
\noindent  
\begin{picture}(80,80) 
\put(0,0){\includegraphics{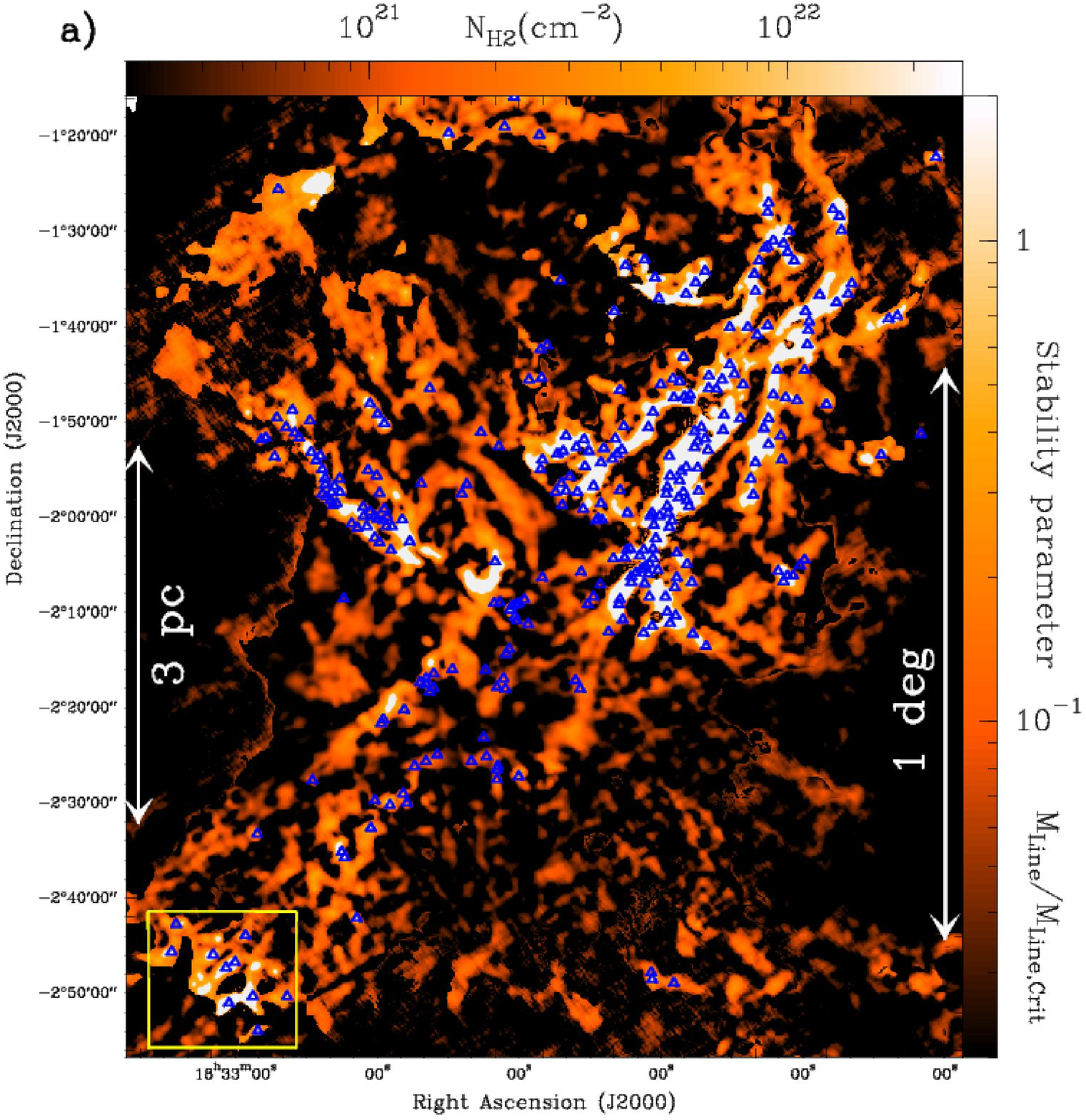}}   
\put(0,0){\includegraphics{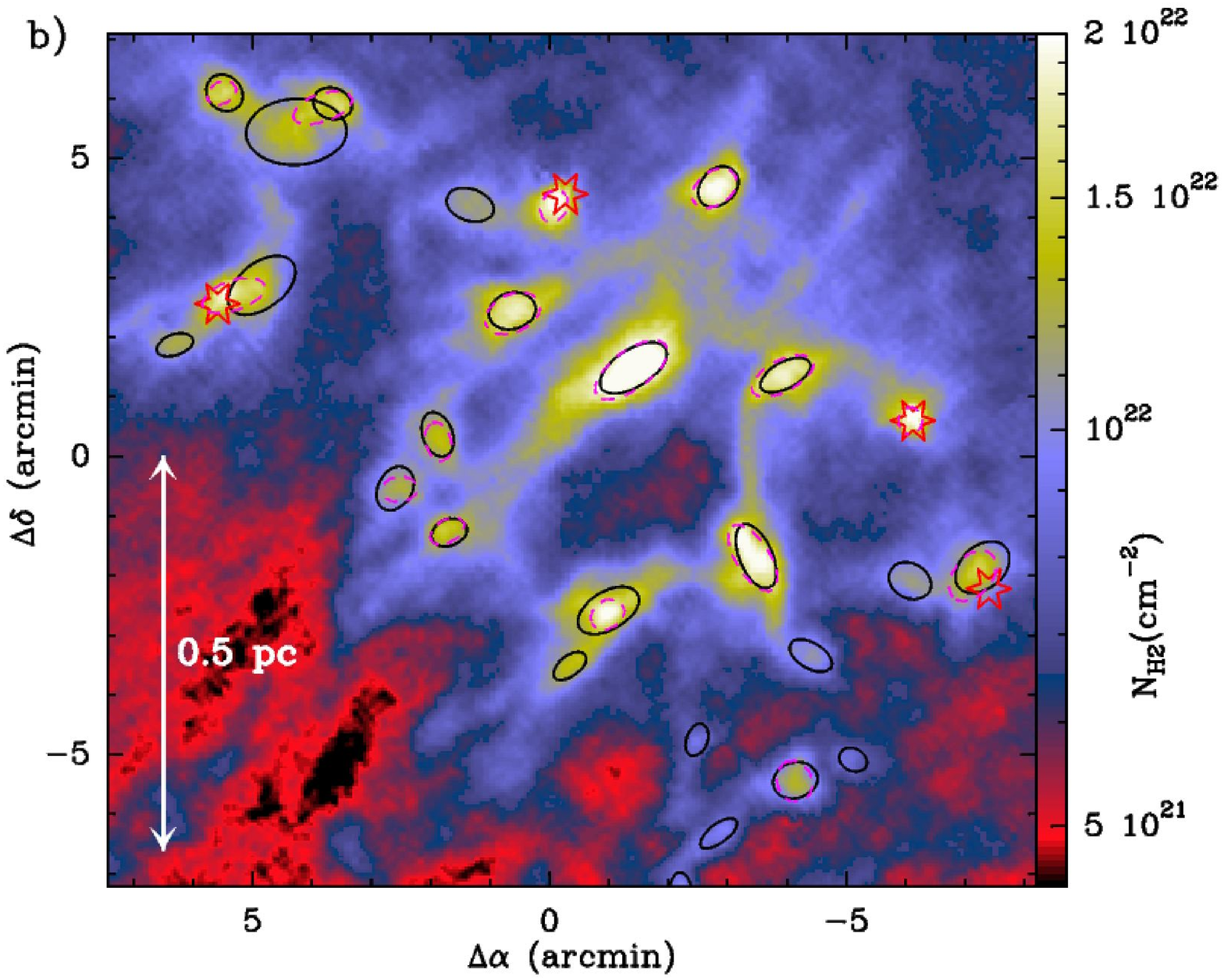}}  
\end{picture} 
\vspace{-0.25cm}
\caption{\small {\bf(a)}  Column density map of a subfield of the Aquila star-forming region derived from $Herschel$ data ({\em Andr\'e et al.,} 2010). 
The contrast of the filaments has been enhanced using a curvelet transform (cf. {\em Starck et al.,} 2003). 
Given the typical width $\sim $~0.1~pc of the filaments  ({\em Arzoumanian et al.,} 2011 -- see Fig.~\ref{histo_width}), this  
map is  equivalent to a {\it map of the mass per unit length along the filaments}. 
The areas where the filaments have a mass per unit length larger than half the critical value $2\, c_s^2/G$ (cf. {\em Inutsuka and Miyama,} 1997 and \S ~5.1) 
and are thus likely gravitationally unstable are highlighted in white. 
The bound prestellar cores identified by {\em K\"onyves et al.} (2010) are shown as small blue triangles.
{\bf(b)}  Close-up column density map  
of the area shown as a yellow box on the left.
The black ellipses mark the major and minor FWHM sizes of the 
prestellar cores  
found with the source extraction algorithm {\it getsources} ({\em Men'shchikov et al.,} 2012); 
four protostellar cores are also shown by red stars. The dashed purple ellipses mark the 
FHWM 
sizes of the sources independently identified with the {\it csar} algorithm ({\em J. Kirk et al.,} 2013). 
The effective  resolution of the  image 
is $\sim 18"$  or $\sim 0.02$~pc at $d \sim 260$~pc. 
}   
\label{Aquila_coldens}   
\end{figure*}

As illustrated in Fig.~\ref{Aquila_coldens} and discussed in detail 
in \S ~3.2 and \S ~6.1 below, another key result from {\it Herschel} 
is the direct connection between filament structure and the formation of cold 
dense cores.  
Filament intersections can provide higher column 
densities at the location of intersection, enhancing star formation even further.  
Indeed, localized mass accumulation due to filament mergers may provide the 
conditions necessary for the onset of clustered star formation.  
Intersecting filaments as a preferred environment for massive star and cluster 
formation have been discussed by {\em Myers} (2009, 2011). 
For example, {\em Peretto 
et al.}  (2012) revealed the ``hub-filament" structure (see {\em Myers} 2009) of 
merged filaments within the B59 core of the Pipe Nebula cloud.  The Pipe 
Nebula is generally not star-forming, with the sole exception of that occurring 
in B59 itself.  Indeed, star formation may have only been possible in the 
otherwise dormant Pipe Nebula due to the increased local density (dominance 
of gravity) in the B59 hub 
resulting from 
filament intersection.  
Merging filaments 
may also influence the formation of higher mass stars.  
For example, merging 
structures may be dominated by a single massive filament that is being fed 
by smaller adjacent 
sub-filaments
(cf. Figs.~\ref{B211_fil_fibers} \& ~\ref{B211_fil}). 
{\em Hill et al.} (2011) and {\em Hennemann et al.} (2012) found examples of such merged 
filaments in the RCW36 and DR21 ridges of Vela C and Cygnus X, respectively.  
At those locations, more massive protostellar candidates or high-mass stars 
are found, possibly since merged filaments have again yielded locally higher 
densities.  
Intersecting filaments may provide the very 
high densities needed locally for cluster formation.  
In the Rosette Molecular Cloud, {\em Schneider et al.} (2012) found a high degree of coincidence 
between high column densities, filament intersections, and the locations of 
embedded infrared clusters throughout that cloud.  

\bigskip
\centerline{\textbf{ 3. DENSE CORE PROPERTIES }} 

\bigskip
\noindent
\textbf{ 3.1 Core definition and core-finding algorithms }
\bigskip

Conceptually, a dense core is an individual fragment or local overdensity which corresponds to a local minimum in the gravitational potential of a molecular cloud. 
A starless core is a dense core with no associated protostellar object.  A prestellar core may be defined as a dense core which is both starless and gravitationally
bound.  In other words, a prestellar core is a 
self-gravitating condensation of gas and dust within a molecular cloud which may potentially form an individual star (or system) by gravitational collapse 
(e.g., {\em Ward-Thompson et al.,} 1994, 2007; {\em Andr\'e et al.,} 2000; {\em Di Francesco et al.,} 2007).  A protostellar core is a dense core within which a protostar has formed.
While in general the gravitational potential cannot be inferred from observations, it turns out to be directly related to the observable column density distribution for
the post-shock, filamentary cloud layers produced by supersonic turbulence in numerical simulations of cloud evolution ({\em Gong and Ostriker,} 2011).  In practical terms, this 
means that one can define a dense core\footnote{In 
mathematical terms, the projection of a dense core onto the plane of  
sky corresponds to a ``descending 2-manifold'' (cf. {\em Sousbie,} 2011) associated to a local 
peak P in column density, i.e., the set of points connected to 
P by integral lines following the gradient of the column density distribution.}
as the immediate vicinity of a local maximum in observed column density maps 
such as those derived from $Herschel$ imaging 
(see Fig.~\ref{Aquila_coldens} for examples).  
One may 
use significant breaks in the gradient of the column density distribution around the core peak to define the core boundaries.
This is clearly a difficult task, however, unless the core is relatively isolated 
and the instrumental noise in the data is negligible (as is often the case with $Herschel$ observations of nearby clouds).

Prestellar cores are observed at the bottom of the hierarchy of interstellar cloud structures and depart from 
the {\em Larson} (1981) self-similar scaling relations (see {\em Heyer et al.,} 2009 for a recent re-evaluation of these relations).
They 
are the smallest units of star formation (e.g., {\em Bergin \& Tafalla,} 2007).  
To first order, known prestellar cores have simple, convex (and not very elongated) shapes, and their density structures approach that 
of Bonnor-Ebert (BE) isothermal spheroids bounded by the external pressure exerted by the 
parent cloud (e.g., 
{\em Johnstone et al.,} 2000; {\em Alves et al.,} 2001; {\em Tafalla et al.,} 2004). 
These BE-like density profiles do not imply that prestellar cores are necessarily in hydrostatic equilibrium, however, and 
are also consistent with dynamic models ({\em Ballesteros-Paredes et al.} 2003).

Since cores are intrinsically dense, they can be identified observationally within molecular clouds as compact objects of submillimeter/millimeter continuum emission (or 
optical/infrared absorption).  In these cases, the higher densities of cores translate into higher column densities than their surroundings, aiding their detection.  Cores 
can also be identified in line emission, using transitions excited by cold, dense conditions, from molecules that do not suffer from freeze-out 
in those conditions, e.g., low-level 
transitions of NH$_{3}$ and N$_{2}$H$^{+}$.  
(Line identification of cores can  
be problematic if the lines used have high optical depths and are self-absorbed, like CO. 
Hyperfine lines of NH$_{3}$ and N$_{2}$H$^{+}$, however, tend to have low optical depths.)
In general, the high mass sensitivities of continuum observations suggest such 
data are the superior means of detecting cores in molecular clouds. Of course, such emission is influenced by both the column density {\it and} the temperature of the 
emitting dust. 
Moreover, 
it can be sometimes difficult to disentangle objects along the line-of-sight (los) without kinematic information, possibly leading to false detections.  
Even with kinematic information, los superpositions of multiple objects can complicate the derivation of reliable core properties 
(cf. {\em Gammie et al.,} 2003). 
Cores can also be difficult to disentangle if the resolution of the observations is too low and cores are blended. 
Most of these problems can be largely mitigated with sufficiently high resolution data; 
in general, being able to resolve down to 
a fraction of 0.1~pc (e.g., $\sim 0.03$~pc or less) appears to be adequate. 

Apart from 
mapping speed, two key advantages of $Herschel$'s broad-band imaging for core surveys  
are that i) dust continuum emission is largely bright and optically thin at far-infrared/submillimeter wavelengths toward cores and thus directly traces well their 
column densities (and temperatures), and ii) the $\sim 18$\arcsec$\,$HPBW 
angular resolution of $Herschel$ at $\lambda = 250\, \mu $m, corresponding to $\sim 0.03$~pc at a distance $d = 350$ pc, 
is sufficient to resolve cores.

In practice, systematic core extraction in wide-field far-infrared or submillmeter dust continuum images 
of highly structured molecular clouds is a complex problem.  
There has been much debate over which method is the 
best for identifying extended sources such as dense cores in submillimeter surveys 
(see, e.g., {\em Pineda et al.,} 2009; {\em Reid et al.,} 2010). 
The problem can be conveniently decomposed into two sub-tasks: 1) source/core detection, and 2) source/core measurement. 
In the presence of noise and background cloud fluctuations, the detection of sources/cores reduces to identifying 
statistically significant intensity/column density peaks based on the information provided by finite-resolution continuum 
images.  The main problem in the measurement of detected sources/cores is finding the spatial extent or ``footprint'' 
of each source/core. 
Previously, submillimeter continuum maps obtained from the ground
were spatially filtered (due to atmosphere) and largely monochromatic, so simpler methods (e.g., the eye {\em Sandell and Knee,} 2001), 
{\it clumpfind} ({\em Williams et al.,}1994; {\em Johnstone et al.,} 2000), and {\it gaussclumps} ({\em Stutzki and G\"usten,} 1990; {\em Motte et al.,} 1998) 
were utilized to identify cores within those data.  

$Herschel$ continuum data require more complex 
approaches, as they are more 
sensitive than previous maps, retaining information on a wider range of scales.  In 
addition, $Herschel$ continuum data can include up to six bands, and have a resolution depending linearly on wavelength.
To meet this challenge, the new {\it getsources} method was devised by {\em Men'shchikov et al.} (2010, 2012) and used to extract
cores in the $Herschel$ Gould Belt survey data.  
Two alternative methods that have also been used on $Herschel$ data include {\it csar} ({\em J. Kirk et al.,} 2013), 
a conservative variant of the well-known segmentation routine {\it clumpfind}, and 
{\it cutex} ({\em Molinari et al.,} 2011), an algorithm that identifies compact sources by analyzing multi-directional 
second derivatives and performing ``curvature'' thresholding in a monochromatic image. 
Comparison of the three methods on the same data shows broad agreement (see, e.g., Fig.~\ref{Aquila_coldens}b), 
with some differences seen in the 
case of closely-packed  groups of cores embedded in a strong background.  Generally speaking, a core can be considered a robust 
detection if it is independently found by more than one algorithm. 
One 
merit of the {\it csar} method is that it can also retain information pertaining to the tree of
hierarchical 
structure 
within 
the field, making it similar to other recent dendrogram 
({\em Rosolowsky et al.,} 2008) or sub-structure codes ({\em Peretto and Fuller,} 2009). 

Once cores have been extracted from the maps, the $Herschel$ observations provide a very sensitive way of distinguishing between 
protostellar cores and starless cores based on the respective presence or absence of point-like 70~$\mu$m emission. 
Flux at 70~$\mu$m can trace very well the internal luminosity of a protostar (e.g., {\em Dunham et al.,} 2008), and $Herschel$ observations 
of nearby ($d < 500$~pc) clouds have the sensitivity to detect even candidate ``first hydrostatic cores'' (cf. {\em Pezzuto et al.,} 2012), the very first and 
lowest-luminosity ($\sim 0.01$--0.1$\, L_\odot $) stage of protostars (e.g., {\em Larson,} 1969; {\em Saigo \& Tomisaka,} 2011; {\em Commer\c con et al.,} 2012 
-- see also chapter by {\em Dunham et al.,}). 

The $Herschel$ continuum data can also be used to divide the sample of starless cores into gravitationally bound and unbound 
objects based on the locations of the cores in a mass vs. size diagram 
(such as the diagram of  Fig.~\ref{growth_diag}a below) 
and comparison of the derived core masses  
with local values of the Jeans or 
BE mass (see {\em K\"onyves et al.,} 2010).

\begin{figure*}   
\setlength{\unitlength}{1mm}  
\noindent  
\begin{picture}(0,0)  
\put(0,0){\includegraphics{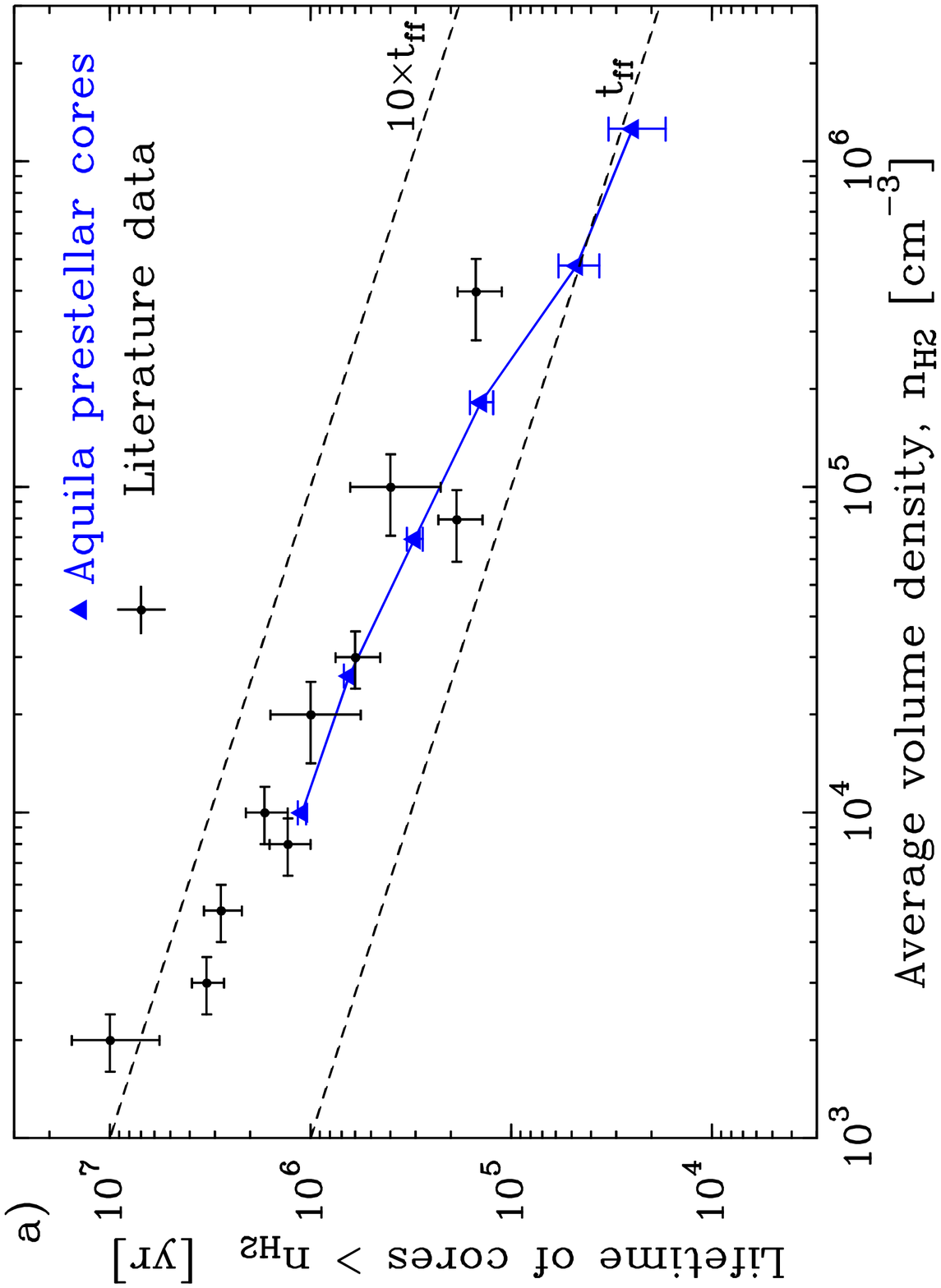}}   
\put(0,0){\includegraphics{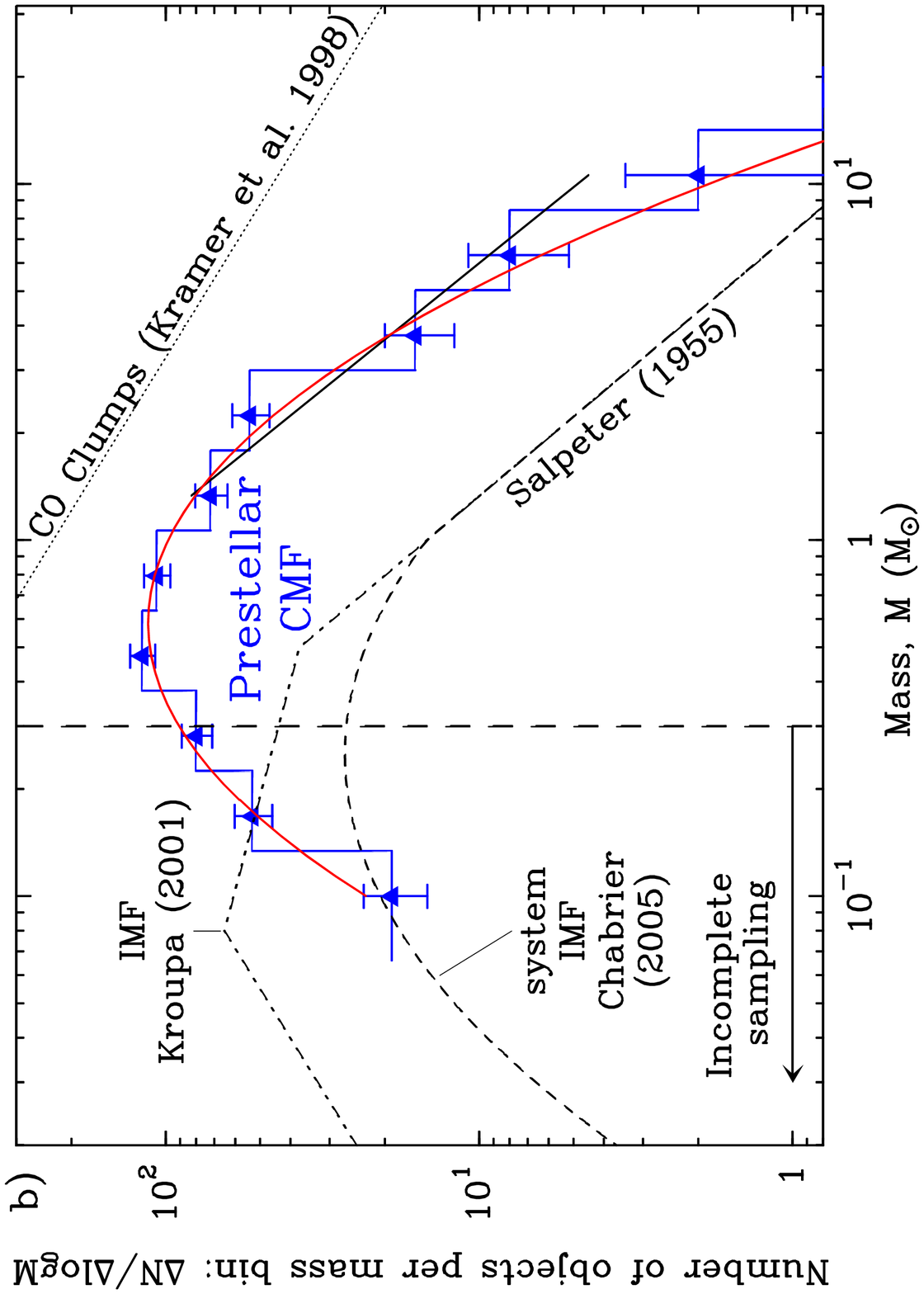}}  
\end{picture}
\vspace{6.75cm}
\caption{\small {\bf(a)}  
Plot of estimated lifetime against minimum average core density (after {\em Jessop and Ward-Thompson,} 2000) 
for the population of prestellar and starless cores identified with $Herschel$ in the Aquila cloud complex (blue triangles -- {\em K\"onyves et al.,} in prep.) 
and literature data (black crosses -- see {\em Ward-Thompson et al.,} 2007 and references therein). 
Note the trend of decreasing core lifetime with 
increasing 
average core density.
For comparison, the dashed lines correspond to the free-fall timescale ($t_{\rm ff}$) and a rough approximation 
of the ambipolar diffusion timescale  ($10 \times t_{\rm ff}$).
{\bf(b)}  Core mass function (blue histogram with error bars) of the $\sim 500$ candidate prestellar cores identified with $Herschel$ 
in Aquila ({\em K\"onyves et al.,} 2010; {\em Andr\'e et al.,} 2010). 
The IMF of single stars (corrected for binaries -- e.g., {\em Kroupa,} 2001),  the IMF of multiple systems (e.g., {\em Chabrier,} 2005), 
and the typical mass spectrum of CO clumps (e.g., {\em Kramer et al.,}1998) are shown for comparison.
A log-normal fit to the observed CMF is superimposed (red curve); it peaks at $\sim 0.6\, M_\odot $, close to the 
Jeans mass within marginally critical filaments at $T \sim 10$~K (see \S ~6.3 below).
}   
\label{Aquila_cmf}   
\end{figure*}   
 
\bigskip
\noindent
\textbf{ 3.2 Spatial distribution of dense cores }
\bigskip

As dense cores emit the bulk of their luminosity at far-infrared and submillimeter wavelengths, 
$Herschel$ mapping observations are ideally suited for taking a deep census of such cold objects in 
nearby molecular cloud complexes. 
Furthermore, since the maps of the $Herschel$ Gould Belt survey (HGBS) essentially cover the entire 
spatial extent of nearby clouds, they provide an unbiased and unprecedented view of the spatial 
distribution of dense cores within the clouds. 
The HGBS results show that more than 70\% of the prestellar cores identified with $Herschel$ are located 
within filaments.  More precisely, bound prestellar cores and deeply embedded protostars (e.g., Class~0 objects) 
are primarily found in filaments 
with column densities $N_{H_2} \simgt 7 \times 10^{21}\, {\rm cm}^{-2} $.
To illustrate, Fig.~\ref{Aquila_coldens} shows the locations of bound cores identified by 
{\em K\"onyves et al.} (2010) on a column density map of the Aquila Rift cloud derived 
from the {\it Herschel} data (see also {\em Andr\'e et al.,} 2010).
As can be plainly seen, 
bound cores are located predominantly in dense filaments (shown in white in Fig.~\ref{Aquila_coldens}a) 
with supercritical\footnote{Throughout this chapter, by supercritical or subcritical filament, we 
mean a filament with $M_{\rm line} > M_{\rm line, crit}$ or $M_{\rm line} < M_{\rm line, crit}$, respectively. 
This notion should not be confused with the concept of a magnetically supercritical or subcritical cloud/core 
(e.g., {\em Mouschovias,} 1991).} 
masses per unit length $M_{\rm line} > M_{\rm line, crit}$, 
where $M_{\rm line, crit} = \frac{2c_{\rm s}^2}{G} $ is the critical line mass of a nearly isothermal 
cylindrical filament (see {\em Inutsuka \& Miyama,} 1997 and \S ~5.1 below) and $c_{\rm s}$ is the isothermal sound speed. 
Interestingly,  the median projected spacing $\sim 0.08$~pc observed between the prestellar cores 
of Aquila roughly matches the characteristic $\sim 0.1$~pc inner width of the filaments (see \S~2.5). 
Overall the spacing between $Herschel$ cores is not periodic, although hints of periodicity 
are observed in a few specific cases (see, e.g., Fig.~\ref{growth_diag}b below).
Only a small ($< 30\% $) fraction of bound cores are found unassociated 
with any filament or only associated with subcritical filaments.  
In the L1641 molecular cloud mapped by the HGBS in Orion~A, {\em Polychroni et al.} (2013) 
report that only 29\% of the prestellar cores lie off the main filaments and that these 
cores tend to be less massive than those found on the main filaments of Orion~A.
The remarkable correspondence between the spatial distribution of compact cores and the 
most prominent filaments (Fig.~\ref{Aquila_coldens}a and {\em Men'shchikov et al.} 2010) suggests 
that {\it prestellar dense cores form primarily by cloud fragmentation along filaments}.
Filaments of significant column density (or mass per unit length) are more likely 
to fragment into self-gravitating cores that form stars.  
We will return to this important point in more detail in \S~6 below.

\bigskip
\noindent
\textbf{ 3.3 Lifetimes of cores and filaments}
\bigskip

Observationally, a rough estimate of the lifetime of starless cores can be obtained from the number ratio 
of cores with and without embedded young stellar objects (YSOs) in a given population (cf. {\em Beichman et al.} 1986), 
assuming a median lifetime of $\sim 2 \times 10^6$~yr for Class~II YSOs (cf. {\em Evans et al.,} 2009). 
Using this technique, {\em Lee and Myers} (1999) found that the typical lifetime of starless 
cores with average volume density $\sim 10^4\ \rm{cm}^{-3}$ is $\sim 10^6$~yr. 
By considering several samples of isolated cores spanning a range of core densities, 
{\em Jessop and Ward-Thompson} (2000) subsequently established that the typical core lifetime decreases as the mean volume 
density in the core sample increases (see also {\em Kirk et al.,} 2005). 

Core statistics derived from recent HGBS studies of nearby star-forming clouds  
such as the Aquila complex are entirely consistent with the pre-$Herschel$ constraints on core lifetimes 
(see {\em Ward-Thompson et al.,} 2007 and references therein).
Figure~\ref{Aquila_cmf}a shows a plot of estimated lifetime versus average volume density, similar to that introduced by  
{\em Jessop and Ward-Thompson} (2000), but for the sample of starless cores identified with $Herschel$ in the Aquila complex 
({\em K\"onyves et al.,} 2010). As can be seen, the typical lifetime of cores denser than $\sim 10^4\ \rm{cm}^{-3}$ 
on average remains $\sim 10^6\, {\rm yr}  $.  
Indeed, all estimated core lifetimes lie between one free-fall time ($t_{\rm ff}$), 
the timescale expected in free-fall collapse, and $10 \times t_{\rm ff}$, roughly the timescale expected for highly subcritical 
cores undergoing ambipolar diffusion (e.g., {\em Mouschovias,} 1991).
Interestingly, Fig.~\ref{Aquila_cmf}a suggests that prestellar cores denser than 
$\sim 10^6\ \rm{cm}^{-3}$ on average evolve essentially on a free-fall timescale. 
Note, however, that statistical estimates of core lifetimes in any given region are quite 
uncertain since it is assumed all observed cores follow the same evolutionary path and that the core/star formation 
rate is constant.  Nevertheless, the fact that similar results are obtained in different regions suggests that the timescales given 
by a plot such as Fig.~\ref{Aquila_cmf}a are at least approximately correct (i.e., within factors of $\sim $ 2--3). 

It is not clear yet whether density is the only parameter controlling a core's lifetime or whether mass also plays an additional role. 
Several studies suggest that massive prestellar cores (i.e., precursors to stars $> 8\, M_\odot $), if they exist at all, are 
extremely rare with lifetimes 
shorter than the free-fall timescale (e.g., {\em Motte et al.,} 2007; {\em Tackenberg et al.,} 2012). 
These findings may be consistent with the trend seen in Fig.~\ref{Aquila_cmf}a and the view that massive stars can only 
form from very dense structures. Alternatively, they may indicate that core evolution is mass dependent (see {\em Hatchell and Fuller,} 2008) 
or that massive protostars acquire the bulk of their mass from much larger scales than a single prestellar core 
(see {\em Peretto et al.,} 2013 and \S ~4.3 below).

The 
above 
lifetime estimates 
for low-mass prestellar cores, coupled with the result that a majority of the cores lie within filaments (\S ~3.2),  
suggest that dense, star-forming filaments such as the B211/B213 filament in Taurus (cf. Fig.~\ref{B211_fil_fibers}) survive for at least $\sim 10^6$~yr. 
A similar lifetime can be inferred for lower-density, non-star-forming filaments (cf. Fig.~\ref{Polaris}) 
from a typical sound crossing time $\simgt 5 \times 10^5$~yr, given their characteristic width $\sim 0.1 $~pc (\S ~2.5) 
and nearly sonic internal velocity dispersion (\SÊ~4.4 and  Fig.~\ref{velo_disp}b below).

\bigskip
\noindent
\textbf{ 3.4 Core mass functions }
\bigskip

Using  {\it getsources},  more than 200 (Class 0 \& Class I) protostars were identified in the $Herschel$ images 
of the whole ($\sim 11$~deg$^2$) Aquila cloud complex ({\em Bontemps et al.,} 2010; {\em Maury et al.,} 2011), along with more than 500 
starless cores $\sim$~0.01--0.1~pc in size (see Fig.~\ref{Aquila_coldens} for some examples). 
On a mass vs.\/ size diagram, most ($> 60\% $) Aquila starless cores lie close to the loci of critical Bonnor-Ebert 
spheres, suggesting that the cores are self-gravitating and prestellar in nature 
({\em K\"onyves et al.,} 2010). 
The core mass function (CMF) derived for the entire sample of $> 500 $ 
starless cores in Aquila is well fit by a log-normal distribution and 
closely resembles the initial mass function (IMF) (Fig.~\ref{Aquila_cmf}b -- {\em K\"onyves et al.,} 2010; {\em Andr\'e et al.,} 2010). 
The similarity between the Aquila CMF and the {\em Chabrier} (2005) system IMF is consistent with an essentially one-to-one 
correspondence between core mass and stellar system mass ($M_{\star \rm sys} = \epsilon_{\rm core}\, M_{\rm core} $).
Comparing the peak of the CMF to the peak of the system IMF suggests that the efficiency $ \epsilon_{\rm core} $ 
of the conversion from core mass to stellar system mass is between $\sim 0.2$ and $\sim 0.4$ in Aquila. 

 \begin{figure*}
 \setlength{\unitlength}{1mm}  
\noindent  
\begin{picture}(60,60)  
\put(0,0){\includegraphics{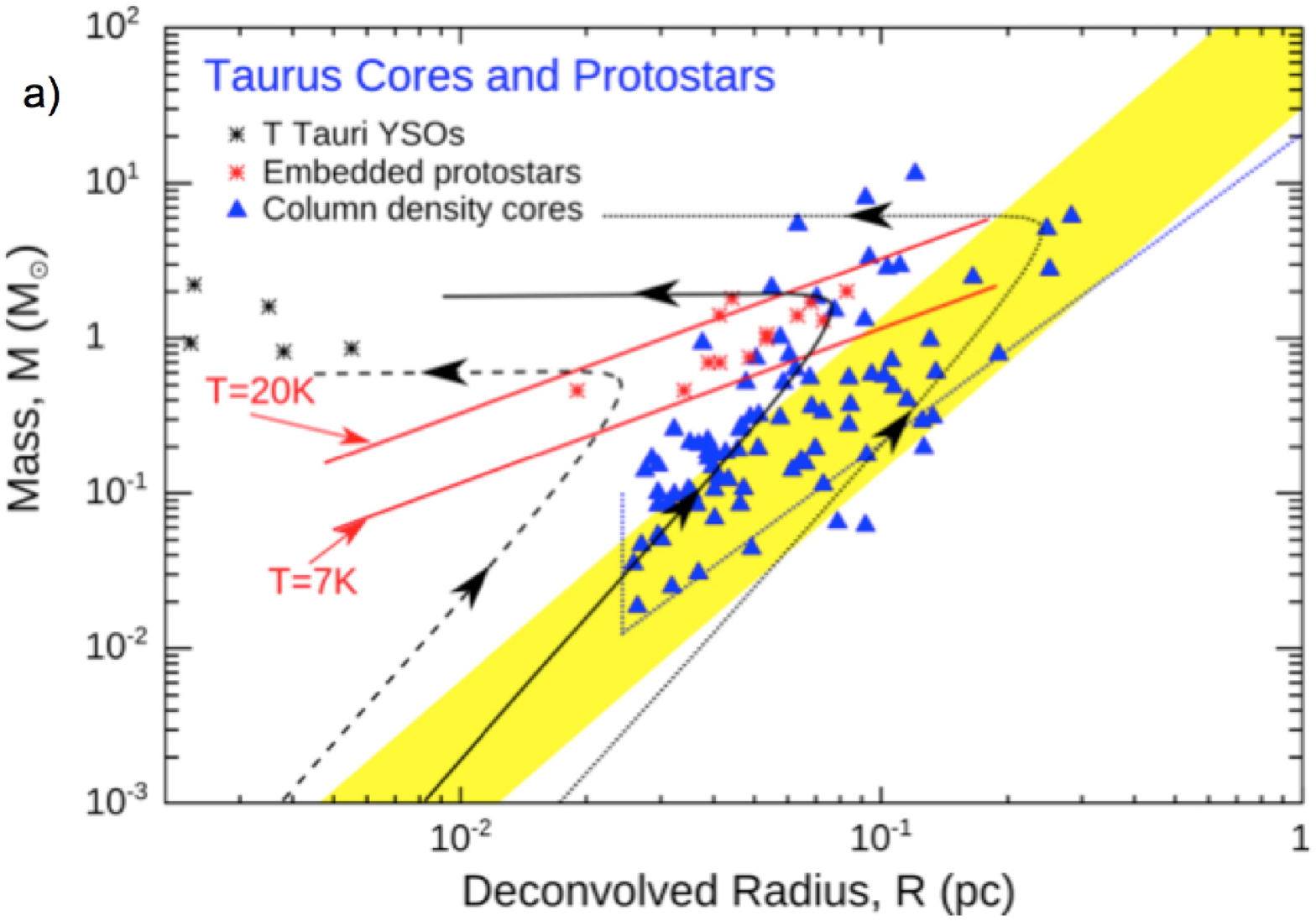}}   
\put(0,0){\includegraphics{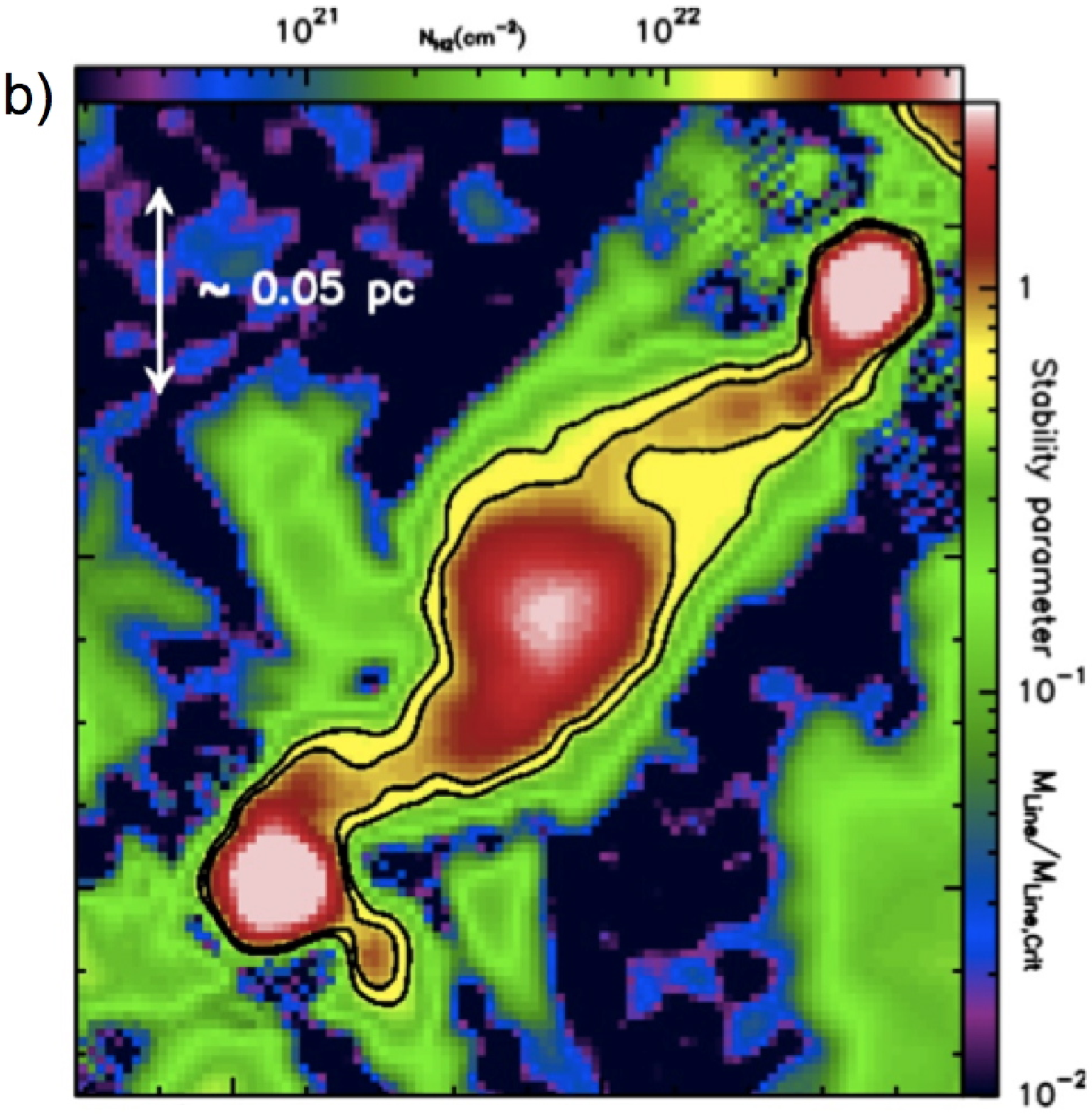}}  
\end{picture}
\vspace{-0.25cm}
 \caption{\small 
 {\bf(a)}  Mass versus size diagram for the cores identified with $Herschel$ in the Taurus S1/S2 region ({\em J. Kirk et al.,} 2013). 
The two red solid lines mark the loci of critically self-gravitating isothermal Bonnor-Ebert
spheres at $T = 7$~K and $T = 20$~K, respectively, while the yellow band shows the region of the mass-size plane 
where low-density CO clumps are observed to lie (cf. {\em Elmegreen and Falgarone,} 1996). 
The blue dotted curve indicates both the mass sensitivity and the resolution limit of the {\em Kirk et al.} (2013) study.
Model evolutionary tracks from {\em Simpson et al.} (2011) are superposed as black lines (dashed, solid, and dotted for
ambient pressures $P_{\rm ext}/k_{\rm B} = 10^6 \, \rm{K\, cm}^{-3} $, $10^5 \, \rm{K\, cm}^{-3} $, and $10^4 \, \rm{K\, cm}^{-3} $, 
respectively), in which starless cores (blue triangles) first grow in mass by accretion of background 
material before becoming gravitationally unstable and collapsing to form protostars (red stars) and T Tauri stars
(black stars). 
{\bf(b)}  Blow-up $Herschel$ column density map of a small portion of the Taurus B211/B213 filament (cf. Fig.~\ref{B211_fil_fibers}a)
 showing two protostellar cores on either side of a starless core ({\em Palmeirim et al.}, in prep.).
 The contrast of filamentary structures has been enhanced using a curvelet transform (see {\em Starck et al.,} 2003).  
 Note the bridges of material connecting the cores along the B211/B213 filament, suggesting that the central starless 
 core may still be growing in mass through filamentary accretion, a mechanism seen in some numerical simulations 
 (cf. {\em Balsara et al.,} 2001 and \S~5).
 }
\label{growth_diag}
\end{figure*}

The first HGBS results on core mass functions therefore confirm the existence of a close similarity between the 
prestellar CMF and the stellar IMF, using a sample $\sim $ 2-9$\times$ larger than those of earlier ground-based studies 
(cf. {\em Motte et al.,} 1998; {\em Johnstone et al.,} 2000; {\em Stanke et al.,} 2006; {\em Alves et al.,} 2007; {\em Enoch et al.,} 2008). 
The efficiency factor $ \epsilon_{\rm core} \sim 30\% $ may be attributable to mass loss due to the effect of outflows during the 
protostellar phase ({\em Matzner and McKee,} 2000). 
More work is needed, however, to derive a reliable prestellar CMF at the low-mass end and fully assess the potential importance of subtle 
observational biases (e.g., background-dependent incompleteness and blending of unresolved groups of cores). 
The 
results from the full HGBS will 
also 
be necessary to characterize fully the nature of the CMF--IMF relationship 
as a function of environment.
Furthermore, as pointed out by, e.g., {\em Ballesteros-Paredes et al.} (2006), establishing a {\it physical} link between the CMF and the IMF is not 
a straightforward task. 

The findings based on early analysis of the $Herschel$ data nevertheless tend to support models of the IMF based on pre-collapse cloud fragmentation 
such as the gravo-turbulent fragmentation picture (e.g., {\em Larson,} 1985;  
{\em Klessen and Burkert,} 2000; {\em Padoan and Nordlund,} 2002; {\em Hennebelle and Chabrier,} 2008). 
Independently of any model, the $Herschel$ observations suggest that one of the keys to the problem of the origin of the IMF  
lies in a good understanding of the formation mechanism of prestellar cores. 
In addition, further processes, such as rotational subfragmentation of prestellar cores into binary/multiple systems during collapse (e.g., {\em Bate 
et al.,} 2003; {\em Goodwin et al.,} 2008) and competitive accretion at the protostellar stage (e.g., {\em Bate and Bonnell,} 2005) may also 
play a role and, e.g.,  
help populate
the low- and high-mass ends of the IMF, respectively (see chapter by {\em Offner et al.}).

\bigskip
\noindent
\textbf{ 3.5 Observational evidence of core growth }
\bigskip

$Herschel$ observations have uncovered a large 
population of faint, unbound starless cores ({\em Ward-Thompson et al.,}  2010; 
{\em Andr\'e et al.,}  2010), not seen in earlier submillimeter continuum data, that 
may be the precursors to the better characterized population 
of prestellar cores. 

As an example, Fig.~\ref{growth_diag}a shows the mass vs.\/ size 
relation for sources found in
the $Herschel$ data of the S1 and S2 regions of Taurus. 
The blue triangles represent the starless and prestellar cores,
red stars are cores with embedded protostars and black stars are
T Tauri stars.
Prior to $Herschel$, 
cores seen in the submillimeter continuum and CO cores 
were found to 
lie in very different areas of the mass--size plane
({\em Motte et al.,}  2001). It was not known if 
this difference was due to selection effects, or whether 
there were two distinct (but possibly related) populations of objects, 
i.e., low-density non-star-forming starless cores 
(cf. {\em Ward-Thompson et al.,}  2010), and higher-density prestellar 
cores ({\em Ward-Thompson et al.,}  1994, 2007).

Figure~\ref{growth_diag} shows that $Herschel$ cores in the 
S1/S2 region of Taurus lie in the region of
mass--size diagram partly over-lapping the CO cores, partly in
the region of the ground-based continuum cores and partly in between
(see also {\em Kirk et al.,} 2013).
This rules out the possibility that there are two unrelated 
populations of objects.  Instead, at least a fraction of CO clumps and 
lower density cores may evolve into prestellar cores by accreting additional 
cloud mass and migrating across the diagram of Fig.~\ref{growth_diag}a, 
before collapsing to form protostars. 
Likewise, {\em Belloche et al.}  (2011) argued that a fraction 
$\sim \, $20--50\% of the 
unbound starless cores  
found in their 
LABOCA observations 
of Chamaeleon 
may be still growing in mass and turn prestellar in the future.  

Given that a majority of 
cores lie within filaments, we speculate 
that this process of core 
growth occurs primarily through filamentary accretion 
(cf. Fig.~\ref{growth_diag}b).  
Interestingly, 
core growth through filamentary accretion 
may have been detected 
toward the starless core L1512 in Taurus ({\em Falgarone et al.,}  2001; see also {\em Schnee et al.,} 2007 for TMC-1C), 
and is  seen in a number of numerical models.
{\em Balsara et al.}  (2001) were amongst 
the first to discuss accretion via filaments and showed that their model was consistent 
with observations of S106.
{\em Smith et al.}  (2011, 2012) described accretion flows from 
filaments onto cores, and {\em G\'omez and V\'azquez-Semadeni}  (2013) discussed
the velocity field of the filaments and their environment, and how
the filaments can refocus the accretion of gas toward embedded cores.

Figure~\ref{growth_diag}a also shows tracks of core evolution in the
mass vs.\/ size diagram proposed by {\em Simpson et al.}  (2011). 
In this picture, cores evolve by accreting mass quasi-statically and 
by maintaining Bonnor-Ebert equilibrium.  In this phase, they move 
diagonally to the 
upper right in the mass vs.\/ size diagram.  Upon reaching Jeans 
instability, a core 
collapses to form a protostar, moving leftward in the diagram.  
However, protostellar cores lie beyond the Jeans limit, in the upper 
left part of the diagram, near the 
resolution limit of the observations 
(they have been given  an 
additional 0.3~M$_\odot$ to account for
mass already accreted onto the protostar).

To substantiate this evolutionary picture, {\em Simpson et al.} (2011) showed that 
cores in L1688 (Ophiuchus) showing weak signs of infall 
(i.e., through blue-asymmetric double-peaked 
line profiles; see {\em Evans,} 1999) appeared close to, or beyond the 
Jeans mass line while cores exhibiting characteristic infall signatures
all appeared beyond it. In L1688,  40\% of the 
cores beyond the Jeans mass line showed signs of infall; 
60\% if those exhibiting possible blue-asymmetric line profiles were included. 

Using this diagram, 
an approximate final stellar mass 
could be derived for any given core by following a Bonnor-Ebert 
track diagonally upward 
to the Jeans mass line.  From that point on in the evolution, an 
efficiency for the collapse 
of the core 
must be assumed to reach an estimate of the final 
stellar mass. If a core continues to accrete whilst collapsing, its 
evolutionary track will 
also move slightly upward after crossing the Jeans mass and turning 
left.  Timescale arguments 
suggest, however, that accretion of background material cannot play 
a major role after the onset 
of protostellar collapse, at least for low-mass protostars 
(cf. {\em Andr\'e et al.,} 2007). 
High-mass protostars may differ significantly from their low-mass 
counterparts since there is 
mounting evidence of parsec-scale filamentary accretion onto 
massive protostellar cores (e.g., {\em Peretto et al.,} 2013).  

\bigskip
\centerline{\textbf{ 4. KINEMATICS OF FILAMENTS AND CORES }}

\bigskip
\noindent
\textbf{ 4.1 Evidence of velocity-coherent structures }
\bigskip

Molecular line observations of several regions have also shown the presence of filamentary structures. 
Single-dish observations of a few regions in the nearby Taurus cloud using molecular lines revealed that 
several filaments are identified in velocity, although in most cases they overlap on the plane of sky
({\em Hacar and Tafalla,} 2011; {\em Hacar et al.,} 2013). 
In 
these cases, the centroid velocity along the filaments does not vary much, suggesting that the 
filaments are coherent (i.e., ``real") structures.
For L1495/B211/B213 in Taurus, the average density was derived from line data, and used to estimate that the 
thickness of the structure along the line of sight is consistent with a cylinder-like structure instead 
of a sheet seen edge-on ({\em Li and Goldsmith,} 2012).
Also, some of these filaments contain dense cores that share the kinematic properties of the associated  
filaments, suggesting that these cores form through filament fragmentation. 

An excellent example is the L1495/B211 filament, observed both in dust continuum by the HGBS ({\em Palmeirim 
et al.,} 2013 -- see Fig.~\ref{B211_fil_fibers}a) and in various lines by {\em Hacar et al.} (2013).  
These data allow for
a direct comparison between properties derived from each tracer.
In the $Herschel$ data, {\em Palmeirim et al.} (2013) identified this filament as a single, thermally supercritical structure $> 5$~pc in length.
In the molecular line data, on the other hand, {\em Hacar et al.} (2013) identified 35 different filamentary components
with typical lengths of $0.5$\,pc, with usually a couple of components overlapping on the plane of sky 
but having distinct velocities.  The masses per unit length (line masses) 
of these components or 
``fibers'' are close to the 
stability value (see \S~5.1 below).  
Interestingly, most of the line-identified fibers can also be seen in the $Herschel$ dust continuum observations  
when large-scale emission is filtered out, enhancing the contrast of small-scale structures in the data (cf. Fig.~\ref{B211_fil_fibers}b). 
The {\em Hacar et al.} (2013) results suggest that some supercritical $Herschel$-identified filaments consist of 
bundles of smaller and more stable fibers.
Other examples of thermal fibers include the narrow ones revealed by high-angular resolution (6\arcsec)
NH$_{3}$ observations of 
the B5 
region in Perseus ({\em Pineda et al.,} 2011).

\begin{figure*}
\setlength{\unitlength}{1mm}  
\noindent  
\begin{picture}(70,70)  
\put(0,0){\includegraphics{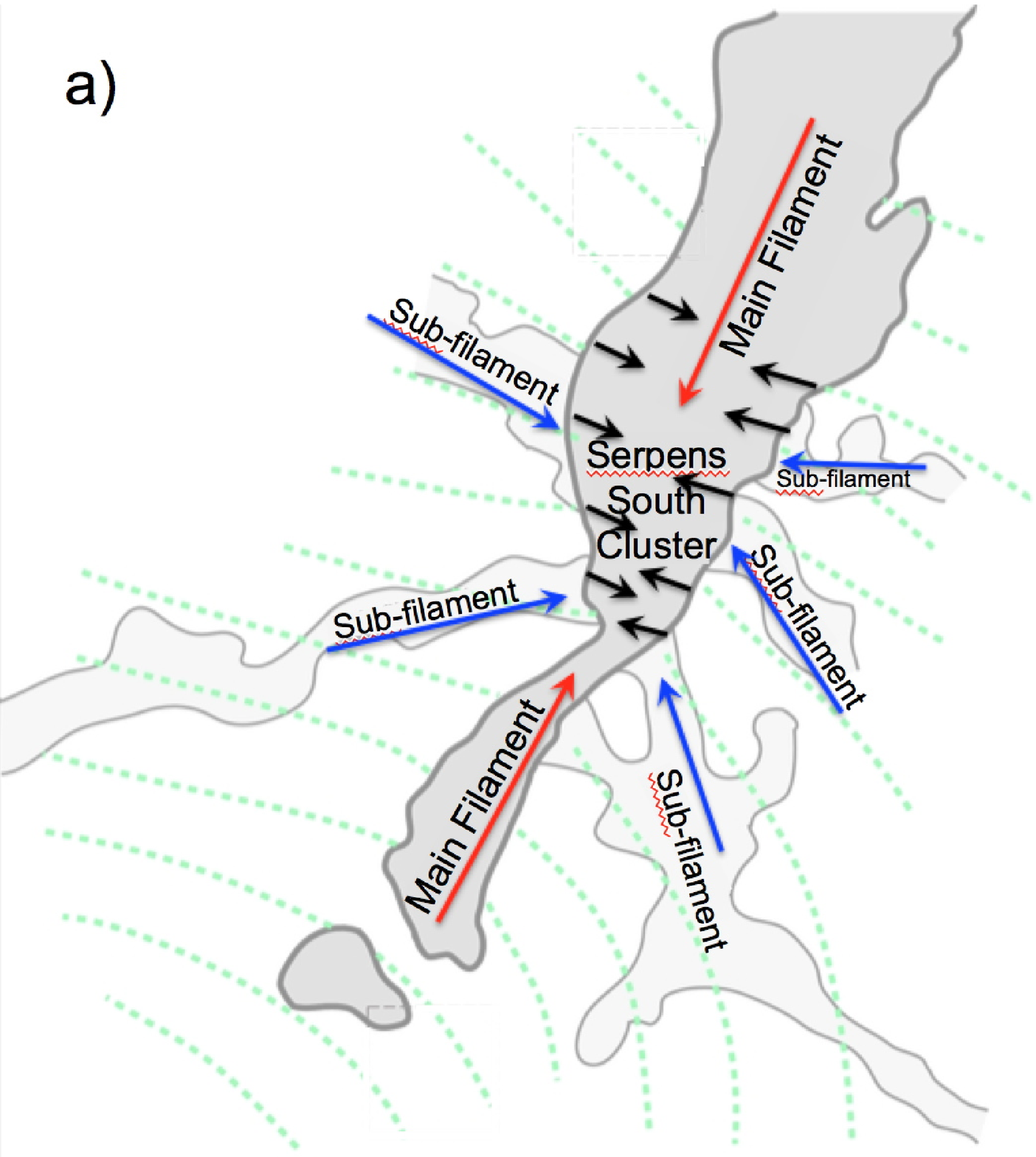}} 
\put(0,0){\includegraphics{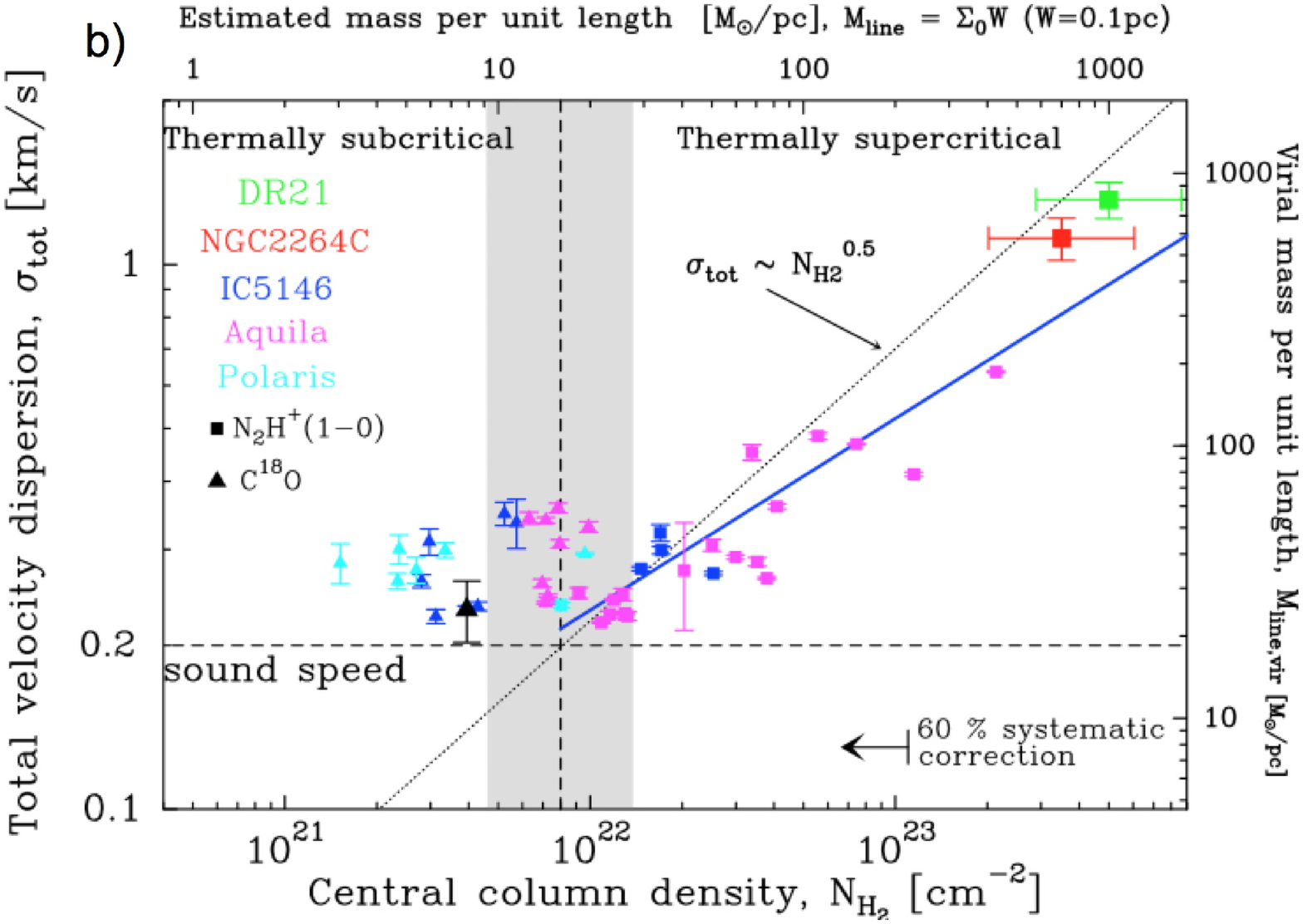}}   
\end{picture} 
\vspace{-0.25cm}
 \caption{\small 
 {\bf(a)}  Sketch of the typical velocity field inferred from line observations (e.g., {\em H. Kirk et al.,} 2013) within and 
 around a supercritical filament (here Serpens-South in the Aquila complex -- adapted from {\em Sugitani et al.,} 2011): 
 Red arrows mark longitudinal infall along the main filament; black arrows indicate radial contraction motions; and 
 blue arrows mark accretion of background cloud material through striations or subfilaments, along magnetic field lines (green dotted lines  from {\em Sugitani et al.}).
{\bf(b)}  
Total (thermal $+$ nonthermal) velocity dispersion versus central  column density for a sample of 
46 filaments in nearby interstellar clouds ({\em Arzoumanian et al.,} 2013).
The horizontal dashed line shows the value of the thermal sound speed $\sim 0.2$ km/s for  $T$~=~10~K. 
The vertical grey band marks the border zone between thermally subcritical and thermally supercritical filaments 
where the observed mass per unit length $ M_{\rm line} $  
is close to the critical value $ M_{\rm line,crit} \sim$ 16~M$_{\odot}$/pc for $T$~=~10~K.
The blue  solid  line shows the best power-law fit $ \sigma_{\rm tot} \propto {N_{\rm H_2}}^{0.36~\pm~0.14}$
to the data points corresponding to  supercritical filaments. 
 }
\label{velo_disp}    
\end{figure*}

\bigskip
\noindent
\textbf{ 4.2 Transition to coherence }
\bigskip

Dense cores in general exhibit velocity dispersions where non-thermal motions (e.g., turbulence) are smaller than thermal 
motions ({\em Myers,} 1983).  
{\em Goodman et al.} (1998) and {\em Caselli et al.} (2002) coined the term ``coherent core,'' to describe the location where
non-thermal motions are subsonic (or at most transonic) and roughly constant.
{\em Goodman et al.} (1998) further showed that the lower density gas around cores, as traced by OH and C$^{18}$O, have 
supersonic velocity dispersions that decrease with size, as expected in a turbulent flow.  Meanwhile, 
NH$_3$ line emission,  tracing dense core gas, 
shows a nearly constant, nearly thermal width.   Therefore, a transition 
must occur at some point  between turbulent gas and more quiescent gas.  

More than a decade later, {\em Pineda et al.} (2010) obtained a large, sensitive NH$_3$ map toward 
B5 in Perseus.  This map showed that the sonic region within B5 is surrounded by NH$_3$ emission displaying supersonic turbulence.  
The transition between subsonic 
and supersonic turbulence (e.g., an increase in the gas velocity dispersion by a factor of 2) occurs in less than a beam width
($<$0.04 pc).  
The typical size of the sonic region is $\sim0.1$\,pc, similar to the width of filaments seen by $Herschel$ (see \S~2.5).

Transitions in turbulence levels have also been observed in other clouds (e.g., L1506 in Taurus; {\em Pagani et al.}, 2010)
and other parts of the Perseus molecular cloud ({\em Pineda et al.}, in prep.).  
Both fields with lower levels of star-formation activity (e.g., L1451 and B5) 
and more active fields (e.g., L1448 and IC348-SW) show well-defined sonic regions completely surrounded by 
more turbulent gas when mapped in NH$_3$ at high resolution.  
Moreover, the transition between sonic and more turbulent gas motions 
is sharp ($<$ a beam width) in all fields, regardless of their environment or level of star-formation activity.

From these results, it is clear that regions of subsonic turbulence mark the locations where material is readily available 
for star-formation.  It is very likely that these sonic regions are formed in the central parts of the filaments identified 
with $Herschel$.  If the sonic regions are massive enough, further substructures, like the isothermal fibers observed in B5,
could develop. 
It should also be stressed that these $\sim 0.1$~pc sonic regions 
are much larger than the milliparsec-scale 
structures in which subsonic interstellar turbulence is believed to  dissipate ({\em Falgarone et al.,} 2009).

\bigskip
\noindent
\textbf{ 4.3 Sub-filaments and low-density striations}
\bigskip

The material within filaments is not static.  In fact, several lines of evidence have shown that: i) material is being 
added to filaments from striations; and ii) filaments serve as highly efficient routes for feeding material into hubs or 
ridges where clustered star formation is ongoing.

In the low-mass  
case, the Taurus cloud presents the most striking evidence for striations along filaments. 
CO maps tracing low-density gas 
show the clear presence of striations in B211/B213 as low-level emission located away from 
the denser filament ({\em Goldsmith et al.,} 2008).  These striations are also prominent 
in $Herschel$ dust 
continuum 
maps of the region from the HGBS  
({\em Palmeirim et al.,} 2013 -- see Figs.~\ref{B211_fil_fibers} 
and  ~\ref{B211_fil}).  
These results suggest that material may be funnelled through the striations onto the main filament.
The typical velocities expected for the infalling material in this scenario are $\sim 0.5$--1\,km\,s$^{-1}$, which are
consistent with the kinematical constraints from the CO observations.

In denser cluster-forming 
clumps like the Serpens South protocluster in Aquila or 
the
DR21
ridge in Cygnus~X, 
filaments are observed joining into a 
central hub where star formation is more active ({\em H. Kirk et al.,} 2013 -- cf. Fig.~\ref{velo_disp}a; {\em Schneider et al.,} 2010). 
Single-dish line observations show classical infall profiles along the filaments, and infall rates of 
$10^{-4}-10^{-3}$\,M$_{\odot}$\,yr$^{-1}$ are derived.  Moreover, velocity gradients 
toward the central 
hub suggest that material is flowing along the filaments,  
at rates similar to the current local star-formation rate.

More recently, ALMA interferometric observations of N$_{2}$H$^{+}$ emission from a more distant IRDC, SDC335, 
hosting high-mass star formation, show the kinematic properties of six filaments 
converging on a central hub ({\em Peretto et al.,} 2013). 
Velocity gradients along these filaments are 
seen, suggesting again that material is being gathered in the central region 
through 
the 
filaments at an estimated rate of $2.5\times10^{-3}$\,M$_{\odot}$\,yr$^{-1}$.  This accretion rate is enough to double 
the mass of the central region in a free-fall time.

\bigskip
\noindent
\textbf{ 4.4 Internal velocity dispersions }
\bigskip

Given the wide range of column densities observed toward $Herschel$-identified filaments (see Sect.~2.5), can
turbulent motions provide effective stability to filaments across the entire range?  
Naively, Larson's relation 
($\sigma_v\propto R^{0.5}$) can give a rough estimate of the level of turbulent motions  
within filaments.  

{\em Heyer et al.} (2009) used the FCRAO Galactic Ring Survey $^{13}$CO (1--0) data to study line widths across 
a wide range of environments.  They focused on the kinematical properties of whole molecular clouds, which 
likely include filaments like those seen with $Herschel$.  They found that the velocity dispersion 
is dependent on the column density, a deviation from Larson's relation.  For example, filaments with higher 
column density have a higher level of turbulent motions 
than is expected from Larson's relation. 
These linewidth estimates, however, were not carried out toward specific filaments, and a more focused 
study on filaments was needed to properly assess the role of nonthermal motions in the dynamical stability of filaments.

Recently, {\em Arzoumanian et al.} (2013) presented molecular line measurements of the internal velocity dispersions 
in 46 $Herschel$-identified filaments.  Figure~\ref{velo_disp}b shows that the level of turbulent motions 
is almost constant 
and does not dominate over thermal support for thermally subcritical and nearly critical filaments.   
On the other hand, a positive correlation is found between the level of turbulent motions and the filament column 
density for thermally supercritical filaments.
This behavior suggests that thermally subcritical filaments are gravitationally unbound with transonic line-of-sight velocity 
dispersions, but thermally supercritical filaments are in approximate virial equipartition (but not necessarily virial equilibrium).

Moreover, it may be possible to glean aspects of filament evolution from plots like Fig.~\ref{velo_disp}b.  For 
example, once a filament becomes thermally supercritical (i.e., self-gravitating) it will contract and 
increase its central column density. Meanwhile, the accretion of material onto the filament (e.g., through striations 
or subfilaments) will increase its internal velocity dispersion and virial mass per unit length.  Following this
line of reasoning, {\em Arzoumanian et al.} (2013) speculated that this accretion process may allow supercritical 
filaments to maintain approximately constant inner widths while evolving.  We will return to this point in \S ~6.5 below.

\bigskip
%
\centerline{\textbf{ 5. THEORETICAL MODELS}}

\bigskip

\noindent
\textbf{ 5.1 Basic physics of self-gravitating 
filaments in the ISM }
\smallskip

Theoretical treatments of filaments in molecular clouds have focused on several different kinds of 
possible filament states: i) equilibria,  ii) collapsing and fragmenting systems that
follow from unstable equilibria, iii) equilibria undergoing considerable radial accretion, and finally
iv)  highly dynamical systems for which equilibrium models are not good descriptions.  Indeed, all of these
aspects of filamentary cloud states may be at play in different clouds or at different times and regions in the same cloud.    
The fact that star forming cores are observed preferentially 
in filaments that are predicted to be gravitationally unstable on the basis of equilibrium
models  (cf. Fig.~\ref{Aquila_coldens}a), 
implies that some aspects of basic equilibrium theory are relevant in real clouds.  We turn first to these results. 

Early papers on filament
structure such as the classic solution for a self-gravitating isothermal
filament ({\em Stodolkiewicz,} 1963; {\em Ostriker,} 1964), assumed that filaments are in cylindrical hydrostatic equilibrium.   

If we multiply Poisson's equation for self-gravity in a cylinder of infinite length 
by $r$ on both sides and integrate from the center 
to the outermost radius $r=R$, we obtain: 
\begin{equation}
       R \left. \frac{d\Phi}{dr} \right|_{r=R} 
     = 2 G \int_{0}^{R}2\pi r \rho dr  
     \equiv 2 G M_{\rm line}, 
\end{equation}
where $\Phi$ denotes gravitational potential and  
$M_{\rm line} $ is 
the mass per unit length (or line mass) defined as 
$M_{\rm line} = \int_{0}^{R}2\pi \rho r dr$. 
This quantity remains constant in a change of the cylinder radius
where $\rho \propto R^{-2}$. 
Thus, the self-gravitational force per unit mass $F_{\rm g,cyl}$
is:  
\begin{equation}
     F_{\rm g,cyl} = \left. \frac{d\Phi}{dr} \right|_{r=R} 
     = 2 \frac{G M_{\rm line}}{R} \propto \frac{1}{R}. 
\end{equation}
On the other hand, if we denote the relation between 
pressure $P$ and density $\rho$ as 
$
     P = K \rho^{\gamma_{\rm eff}}, 
$
the pressure gradient force per unit mass 
$F_{\rm p}$ scales as:  
\begin{equation}
     F_{\rm p} = \frac{1}{\rho}
                 \frac{\partial P}{\partial r}
               \propto  
                 R^{1-2\gamma_{\rm eff}}. 
                                         \label{eq:GradP}
\end{equation}
Therefore, we have:  
\begin{equation}
     \frac{ F_{\rm p} }{ F_{\rm g,cyl} } 
     \propto R^{2-2\gamma_{\rm eff}}. 
\end{equation}
If $\gamma_{\rm eff} > 1$, 
the pressure gradient force will dominate self-gravity for a sufficiently small 
outer radius $R$. 
In contrast, if $\gamma_{\rm eff} \leq 1$, 
radial collapse will continue indefinitely 
once self-gravity starts to dominate over the pressure force. 
Therefore, we can define the critical ratio of specific heats for the radial stability of 
a self-gravitating cylinder as 
$
        \gamma_{\rm crit,cyl} = 1. 
$
In the case of $\gamma_{\rm eff}~=~1$, and under a sufficiently small 
external pressure, a cylinder will be in hydrostatic equilibrium
only when its mass per unit length has the special value for which $F_{\rm p}=F_{\rm g,cylinder}$. 
Therefore, we can define a critical mass per unit length for an 
isothermal cylinder: 
\begin{equation}
     M_{\rm line, crit} \equiv \int_0^{\infty} 2\pi \rho_4(r)rdr 
     =  \frac{2c_{\rm s}^2}{G} , 
\end{equation}
where $ \rho_4(r) $ is given by Eq.~(1) in Sect.~2.4 with $p =4$.
Note that, when $\gamma_{\rm eff}$ is slightly less than 1 (as observed -- cf. Fig.~\ref{B211_prof}b), 
the maximum line mass 
of {\it stable} cylinders is 
close to the critical value $M_{\rm line, crit}$ for an isothermal cylinder.

Using similar arguments, 
the critical $\gamma_{\rm eff}$ is found to be $\gamma_{\rm crit,sphere} = 4/3$ 
for a sphere  and $\gamma_{\rm crit,sheet} = 0$ for a sheet. 
The thermodynamical behavior of molecular clouds 
corresponds to $\gamma_{\rm eff} \simlt 1$ (e.g., {\em Koyama and Inutsuka,} 2000 --
see also Fig.~\ref{B211_prof}b for the related anti-correlation between dust 
temperature and column  density). 
Therefore, the significance of filamentary geometry 
can be understood in terms of ISM thermodynamics. 
For a sheet-like cloud, there is always an equilibrium configuration since the internal pressure gradient can always become strong 
enough to halt the gravitational collapse 
independently of its initial state (e.g., {\em Miyama et al.,} 1987; {\em Inutsuka and Miyama,} 1997). 
In contrast, the radial collapse of an isothermal cylindrical cloud cannot be halted and no equilibrium is possible
when the mass per unit length exceeds the critical  value $M_{\rm line, crit}$. 
Conversely, if the mass per unit length of the filamentary cloud 
is less than $M_{\rm line, crit}$, gravity can never be made to dominate by increasing the external pressure, 
so that the collapse is always halted at some finite cylindrical radius.   

Another way of describing this behavior 
is in terms of the effective 
gravitational energy per unit length of a filament, 
denoted $\mathcal{W} $ which takes a simple
form ({\em Fiege and Pudritz,} 2000):
\begin{equation}
\mathcal{W} = - \int \rho r  \frac{d\Phi}{dr}  d\mathcal{V} = -M_{\rm line}^2 G,
\end{equation}
where $\mathcal{V} $ is the volume per unit length (i.e., the cross-sectional area) 
of the filament. $\mathcal{W} $ is the gravitational term in the virial equation.
It is remarkable in that it remains constant during any radial contraction caused by an increased external 
pressure. 
The scaling was first derived by {\em McCrea} (1957).
If a filament is initially in equilibrium, the 
effective 
gravitational energy can never be made to dominate over the pressure
support by squeezing it ({\em McCrea,} 1957; {\em Fiege and Pudritz,} 2000).  
Thus, filaments differ markedly from isothermal spherical clouds which can always be induced to collapse by 
a sufficient increase in external pressure (e.g., {\em Bonnor,} 1956; {\em Shu,} 1977). 

The critical mass per unit length $M_{\rm line, crit}  \approx 16\, M_\odot {\rm pc}^{-1}$ $\times (T_{\rm gas}/10\, {\rm K}) $
as originally derived, depends only on gas temperature $T_{\rm gas}$. 
This expression can be readily generalized to include the presence of non-thermal gas motions.
In this case, the critical mass per unit length becomes 
$M_{\rm line, vir} = 2\, \sigma_{\rm tot}^2/G $, 
also called the virial mass per unit length, where $\sigma_{\rm tot} = \sqrt{c_s^2 + \sigma_{\rm NT}^2} $ 
is the total one-dimensional velocity dispersion including both thermal and non-thermal components ({\em Fiege and Pudritz,} 2000). 
Clearly, both the equation of state of the gas and filament turbulence will play a role in 
deciding this critical line mass.

\begin{figure*}   
\epsscale{2.1}
\plotone{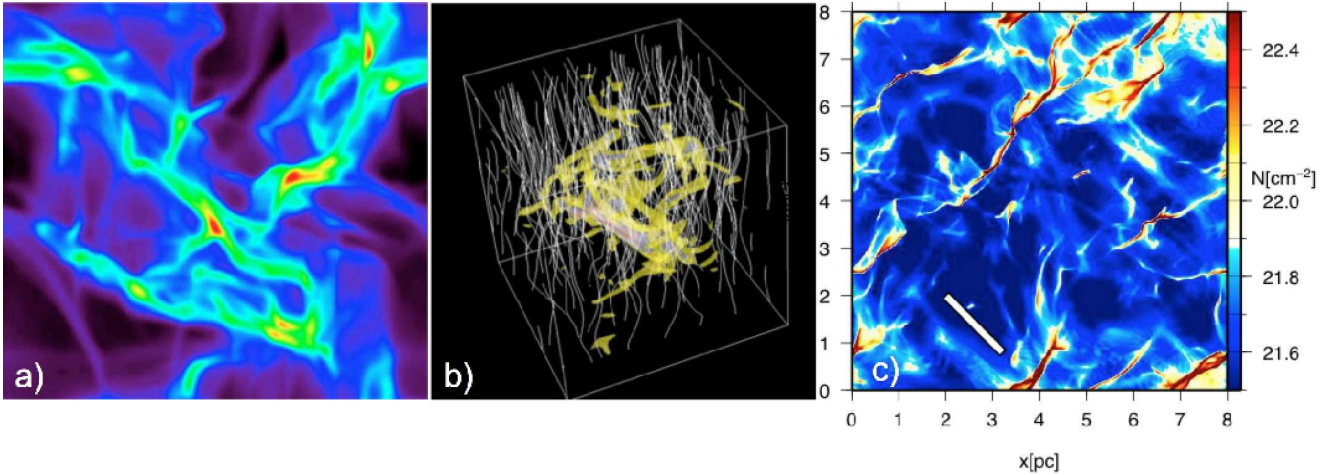}
\caption{\small 
{\bf(a)}  Formation of filaments and cores from shock compression in numerical simulations of isothermal supersonic turbulence 
 with rms Mach number $\sim 10$ ({\em Padoan et al.,} 2001).
{\bf(b)}  Three-dimensional view of flattened filaments resulting from mass accumulation along magnetic field lines 
 in the MHD model of {\em Li et al.} (2010). The 
 magnetic field favors the formation of both low- and high-mass stars 
 compared to intermediate-mass stars, due to fragmentation of high-density filaments and global clump collapse, respectively.
{\bf(c)}  Face-on column density view of a shock-compressed dense layer in 
numerical MHD simulations 
by {\em Inutsuka, Inoue, and Iwasaki,} in preparation.
 The color scale (in cm$^{-2}$) is shown on the right. 
 The mean magnetic field is in the plane of the layer and its direction is shown by a white bar.
 Note the formation of dense, magnetized filaments whose axes are almost perpendicular to the mean magnetic field. 
 Fainter ``striation''-like filaments can also be seen, which are almost perpendicular to the dense filaments. 
 }   
\label{Simulations1}   
\end{figure*}   

Magnetic fields also play an important role in filamentary equilibria.  The most general magnetic geometry consists of both poloidal 
as well as toroidal components.  The latter component  arises quite generally in oblique shocks as well as in shear flows, both
of which  impart a sense of 
local spin to the gas and twisting of the poloidal field lines  (see chapter by {\em H.-b. Li et al.}). 
Toroidal fields also 
are generated during the collapse of magnetized cores (e.g., {\em Mouschovias and Paleologou,} 1979) 
and  carry off substantial fluxes
of angular momentum  (see chapter by {\em Z.Y. Li et al.}). 
The general treatment of the virial theorem for 
magnetized gas  
({\em Fiege and Pudritz,} 2000) shows that  the poloidal
field that threads a filament provides a pressure that helps support a filament against gravity.  
On the other hand, the toroidal field contribution to the 
total magnetic energy is destabilizing; it squeezes down on the surface of the filament and assists gravity.  
Thus the total magnetic energy can 
be negative in regions where the toroidal field dominates,  
and positive where the poloidal field dominates.
{\em Fiege and Pudritz} (2000) 
showed that  $M_{\rm line, vir}$ is modified in the case of magnetized filaments, i.e., 
$M_{\rm line, vir}^{\rm mag} = M_{\rm line, vir}^{\rm unmag} \times  \left(1 - \mathcal{M}/|\mathcal{W}| \right)^{-1}$, 
where $\mathcal{M} $ is the magnetic energy per unit length 
(positive for poloidal fields and negative for toroidal fields). 
While molecular gas in dense regions typically has $|\mathcal{M}|/|\mathcal{W}| \simlt 1/2 $ 
({\em Crutcher,} 1999, 2012),  there are large variations over these measurements.  For the typical case,  
$M_{\rm line, vir}^{\rm mag} $ differs from $M_{\rm line, vir}^{\rm unmag} \equiv  2\, \sigma_{\rm tot}^2/G $ by less than a factor of 2. 
Radial profiles for toroidal isothermal models follow an $\rho \propto r^{-2}$ form, 
in agreement with observations (cf. \S ~2.4).
 
The essence of a virial theorem analysis of filaments can be demonstrated in a diagram  
where the ratio of observed
surface pressure to total turbulent pressure 
($P_S/$$<$$P$$>$)  of a filament is plotted against 
the ratio of its observed mass per unit length to the critical value ($ M_{\rm line}/M_{\rm line, vir} $) (see {\em Fiege and Pudritz,} 2000).
The virial theorem for filaments gives:\\  
$P_S/$$<$$P$$> = 1 - \left[ (M_{\rm line}/M_{\rm line, vir})  \times  (1 - ( \mathcal{M}/|\mathcal{W}| ) )\right]$.\\
In this diagram,  purely hydrodynamic filaments can be easily discriminated from filaments with net positive or negative magnetic energy.
Analysis by {\em Contreras et al.} (2013) of a collection of filaments in the ATLASGAL catalogue of $870\, \mu$m images as well as $^{13}$CO data
revealed that most of the filaments fell in the 
part of this virial diagram where toroidal fields dominate (see Fig.~3 of {\em Contreras et al.} 2013).   
In addition, work by {\em Hernandez and Tan} (2011) on several IRDC filaments indicated the presence of large surface pressures 
($P_S/$$<$$P$$>) > 1$, 
suggesting that these systems may be quite out of equilibrium.  

Even moderate fields can have a very strong effect on the formation of unstable regions in filaments ({\em Tilley and Pudritz,} 2007; {\em Li et al.,} 2010).
Figure~\ref{Simulations1}b shows that rather flattened filaments form in 3D MHD simulations because the magnetic field strongly affects 
the turbulent flow ({\em Li et al.,} 2010),  forcing material to flow and accumulate along field lines.   
The magnetic Jeans mass,  $M_J = (B_o / 2 \pi G^{1/2} \rho_o^{2/3})^3 $,
exceeds the thermal Jeans mass in this case, 
suppressing the formation of intermediate-mass stars but leaving the more massive stars relatively unaffected.  
 
The fragmentation properties of filaments and sheets differ from those of spheroidal clouds in that there is 
a preferred scale for gravitational fragmentation which directly scales with the scale height of the filamentary or sheet-like medium 
(e.g., {\em Larson,} 1985; see also \S ~5.3).
In the spherical case, the largest possible 
mode (i.e., overall collapse of the medium) has the fastest growth rate 
so that global collapse tends to overwhelm the local collapse of finite-sized density perturbations,  
and fragmentation is generally suppressed in the absence of sufficiently large initial density enhancements (e.g., {\em Tohline,} 1982). 
It is also well known that spherical collapse quickly becomes strongly centrally concentrated ({\em Larson,} 1969; {\em Shu,} 1977), 
which tends to produce a single central density peak rather than several condensations (e.g., {\em Whitworth et al.,} 1996). 
In contrast, sheets have a natural tendency to fragment into filaments (e.g., {\em Miyama et al.,} 1987) and filaments with masses per unit length 
close to $M_{\rm line, crit}$ a natural tendency to fragment into spheroidal cores (e.g., {\em Inutsuka and Miyama,} 1997). 
The filamentary geometry is thus the most favorable configuration for small-scale perturbations to collapse locally and 
grow significantly before global collapse overwhelms them ({\em Pon et al.,} 2011, 2012; {\em Toal\'a et al.,} 2012). 
To summarize, theoretical considerations alone emphasize the potential importance of filaments for core and 
star formation.

\bigskip
\noindent
\textbf{ 5.2 Origin of filaments: various formation models }
\bigskip

Since the mid 1990s, simulations of supersonic turbulence have consistently shown that gas is rapidly compressed 
into a hierarchy of 
sheets and filaments (e.g., {\em Porter et al.,} 1994; {\em V\'azquez-Semadeni,} 1994; {\em Padoan et al.,} 2001). 
Furthermore, when gravity is added into turbulence simulations,  the denser gas undergoes gravitational collapse to form stars 
(e.g., {\em Ostriker et al.,} 1999; 
{\em Ballesteros-Paredes et al.,} 1999; 
{\em Klessen and Burkert,} 2000; 
{\em Bonnell et al.,} 2003; 
{\em MacLow and Klessen,} 2004; {\em Tilley and Pudritz,} 2004; {\em Krumholz et al.,}  2007). 
There are many sources of supersonic turbulent motions in the ISM out of which molecular clouds can arise, i.e., 
galactic spiral shocks in which most giant molecular clouds form, supernovae, stellar winds from massive stars, 
expanding HII regions, radiation pressure, cosmic ray streaming,
Kelvin-Helmholtz and Rayleigh-Taylor instabilities, gravitational instabilities, and bipolar 
outflows from regions of star formation ({\em Elmegreen and Scalo,} 2004).  An important aspect of 
supersonic shocks is that they are highly dissipative. Without constant replenishment, 
they damp within a cloud crossing time.  

Filaments are readily  produced 
in the complete absence of gravity.  Simulations of turbulence often
employ a spectrum of plane waves that are random in direction and phase.  As is well known, the crossing of 
two planar shock wave fronts is a line - the filament (e.g., {\em Pudritz and Kevlahan,} 2013).    
The velocity field in the vicinity
of such a filament shows a converging flow perpendicular to the filament axis as well as 
turbulent vortices in the filament wake.  
{\em Hennebelle} (2013) reported that filaments can be formed by the velocity shear that is ubiquitous in magnetized turbulent media 
(see also {\em Passot et. al.,} 1995). 
Others have attributed filaments to being 
stagnation regions in turbulent media, where filaments are considered to be rather transient structures (e.g., {\em Padoan et al.,} 2001). 
{\em Li et al.} (2010) have shown that filaments are formed preferentially perpendicular to the magnetic field lines in strongly magnetized turbulent clouds 
(see also {\em Basu et al.,} 2009). 
An important and observationally desirable aspect of supersonic turbulence 
is that it produces  hierarchical structure that can be described by a lognormal
density distribution ({\em V\'azquez-Semadeni,} 1994).  
A lognormal arises whenever the probability
of each new density increment in the turbulence is independent of the
previous one.  In a shocked medium, the density at any point is the product of the 
shock-induced, density jumps and multiplicative processes of this
kind rapidly converge to produce lognormals  as predicted by the central limit theorem
({\em Kevlahan and Pudritz,} 2009). 

As the mass accumulates in a filament, it becomes gravitationally unstable and filamentary flows parallel to the filament
axis take place as material flows toward a local density enhancement (e.g., {\em Banerjee et al.,} 2006).   Since the advent 
of sink particles to trace subregions which collapse to form stars, simulations have consistently shown that there is a close
association between sink particles and filaments (e.g., {\em Bate et al.,} 2003).  
The mechanism
of the formation of such fragments in terms of gravitational instabilities above a critical line mass does not seem
to have been strongly emphasized  in the simulation literature until recently, however.

Simulations show that as gravity starts to become more significant, it could be as  important as turbulence in 
creating filamentary web structures and their associated ``turbulent'' velocity fields in denser collapsing regions of clouds. 
Simulations of colliding gas streams show that the resulting dense clouds are always far from equilibrium even though a 
simplistic interpretation of the virial theorem might indicate otherwise.   The point is that as clouds build up, 
the global gravitational self energy of the system $\vert E_g \vert $ grows to the point where gravitational collapse begins.
The kinetic energy resulting from the collapse motions $E_k $ tracks this gravitational term in such a way that  $ \vert E_g \vert \simeq 2 E_k$
(e.g.,  {\em V\'azquez-Semadeni et. al.,} 2007;  {\em Ballesteros-Paredes et.al.,} 2011), mimicking virial equilibrium.  
Thus, gravitational collapse in 
media that have a number of Jeans masses will have multiple centers of collapse, and the velocity fields that are set up 
could be interpreted as turbulence (see {\em Tilley and Pudritz,} 2007,  for magnetic analogue).  Such systems are never near equilibrium
however.  

As noted, simulations of colliding streams produce clouds that are confined to a flattened layer.  We therefore
consider the fragmentation of a dense shell created by 
1D compression  (e.g., by an expanding HII region, 
an old supernova remnant, or collision of two clouds) in more detail -  specifically  
the formation of filaments by self-gravitational fragmentation of sheet-like clouds. 
As discussed in \S ~5.1,  
since the critical ratio of specific heats $\gamma_{\rm crit,sheet}= 0$, 
sheet-like configurations are stable against compression.  
Here, the compressing motions are in the direction of the thickness
and these can always be halted by 
the resulting increase in (central) pressure of the sheet, justifying the analysis of quasi-equilibrium sheet-like clouds. 

Self-gravitational instability of a sheet-like equilibrium is well known. 
The critical wavelength for 
linear perturbations on the sheet is 
a few times its thickness.  
Sheet-like clouds are unstable to perturbations whose wavelengths are larger than the critical wavelength, and the most unstable wavelength is about twice the critical wavelength.  
The growth timescale is on the order of the free-fall timescale. 
The detailed analysis by {\em Miyama et al.} (1987) on the non-linear growth of unstable perturbations 
shows that the aspect ratio of dense regions increases with time, resulting 
in the formation of filaments. 
Thus, sheet-like clouds are expected to break up into filaments whose separations are about twice the critical wavelength (i.e., several times the thickness of the sheet). 
If the external pressure is much smaller than the 
central pressure,  
the thickness of the sheet is on the order of the Jeans length 
($\lambda_{\rm J} \sim c_{\rm s} t_{\rm ff}$).  
In contrast, if the external pressure is comparable to the central pressure, the thickness of the sheet can be much smaller than the Jeans length. 
As pointed out by {\em Myers} (2009), the fragmentation properties expected for compressed sheet-like clouds 
resemble the frequently observed ``hub-filament systems'' (see also {\em Burkert and Hartmann,} 2004).
 
Magnetic fields that are perpendicular to the sheet tend to stabilize the sheet. 
Indeed, 
if the field strength  is larger than the critical value ($B_{\rm crit} = 4 \pi^2 G \Sigma^2$, where $\Sigma$ is the surface density of the sheet), 
the sheet is stable against gravitational fragmentation ({\em Nakano and Nakamura,} 1978). 
In contrast, if the magnetic field is in the plane of the sheet, the stabilizing effect is limited, but the direction of the field determines the directions of fragmentation ({\em Nagai et al.,} 1998).   
If the external pressure is much smaller than the central pressure of the sheet, the sheet will fragment into filaments whose axis directions are perpendicular to the mean direction of magnetic field lines.  
In this case, the masses per unit length of the resulting filaments are expected to be about twice the critical value 
($M_{\rm line} \sim 2 M_{\rm line, crit}$). 
Such filaments may be the dense filaments observed with $Herschel$.  
On the other hand, if the external pressure is comparable to the central pressure in the sheet, the resulting filaments will be parallel to the mean magnetic field lines. 
In this case, the masses per unit length of the resulting filaments are smaller than the critical value 
($M_{\rm line} < M_{\rm line, crit}$). 
Such filaments may be the ``striations'' seen with $Herschel$.  

Numerical simulations of the formation of molecular clouds through 1D compression
are now providing interesting features 
of more realistic evolution ({\em Inoue and Inutsuka,} 2008, 2009, 2012).
Figure~\ref{Simulations1}c 
shows a snapshot of the face-on view of a non-uniform molecular cloud compressed by a shock wave travelling at 10 km/s ({\em Inutsuka et al.,} in prep.). 
The magnetic field lines are mainly in the dense sheet of 
compressed gas. 
Many dense filaments are created, with axes perpendicular to the mean magnetic field lines. 
We can also see many faint filamentary structures that mimic observed ``striations'' and are almost parallel to the mean magnetic field lines. 
Those faint filaments are feeding gas onto the dense filaments just as envisioned in \S ~4.4.

Although further analysis is required to give a conclusive interpretation of the observations, a simple picture 
based on 1D compression of a molecular cloud may 
be a promising direction for understanding sheet and 
filament formation. 

\bigskip
\noindent
\textbf{ 5.3 Fate of dense filaments }
\bigskip

As shown earlier in 
Sect.~3.2 and Fig.~\ref{Aquila_coldens}a, 
dense filaments with supercritical masses per unit length harbor star forming cores.  
Here, we theoretically outline what 
happens in such dense filaments. 

\begin{figure}[Htb]
  \epsscale{0.83}
  \plotone{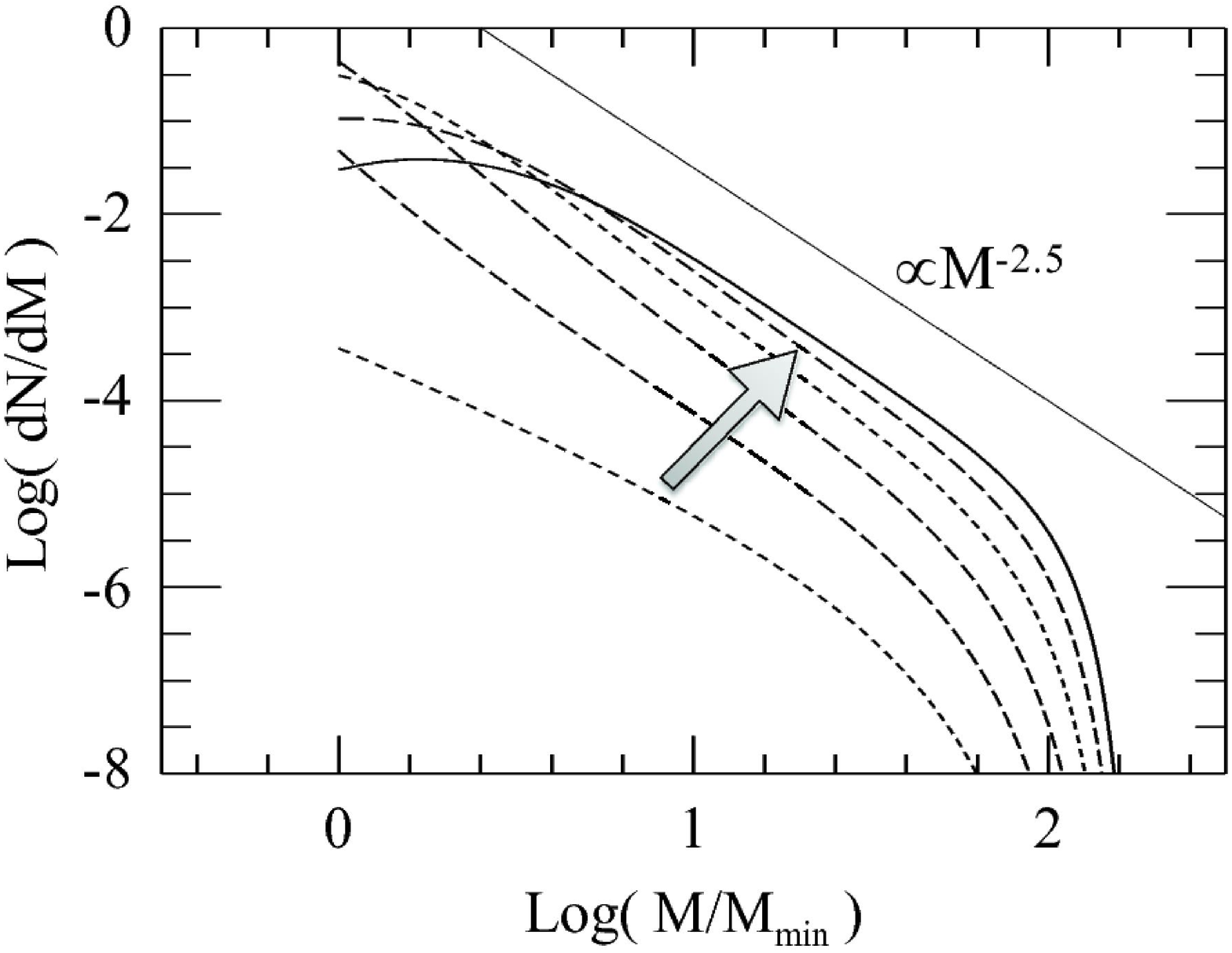}
 \vspace{-0.3cm}
 \caption{\small 
         Evolution of the mass function of dense cores produced by filament fragmentation 
         in the case of $\delta(k)\propto k^{-1.5}$ ({\em Inutsuka,} 2001).
         The curves correspond to the snapshots at $t/t_{\rm g} =$ 0 (dashed), 2, 4, 6, 8, and 10 (solid), where $t_{\rm g}$ is the free-fall timescale, and the arrow indicates the time sequence. 
         $M_{\rm min}$ corresponds to the most unstable wavelength of the quasi-static filament. 
         The thin straight line corresponds to $dN/dM \propto M^{-2.5}$. 
         }
\label{fig:PSMF}  
\end{figure}

Self-gravitating cylindrical structures are unstable to perturbations whose wavelengths are sufficiently large. 
For simplicity, we first focus on unmagnetized equilibrium filaments with isothermal or polytropic equation of state ({\em Inutsuka and Miyama,}  1992). 
The general behavior of the growth rate does not depend much on the polytropic index. 
A cylinder is unstable for wavelengths larger than about twice its diameter, and the most unstable wavelength is about 4 times its diameter.  
The growth timescale is somewhat longer than the free-fall timescale.  
{\em Nagasawa}  (1987) and {\em Fiege and Pudritz}  (2000) investigated the effect of an axial magnetic field on the stability of the isothermal equilibrium filament. 
The stabilizing effect of an axial field is strong in the case where the line-mass of the filament is much less than the critical value, 
while the effect saturates and does not qualitatively change the character of the instability for a filament of nearly critical line-mass. 
An axial field slightly increases the critical line mass and the most unstable wavelength. 
In contrast, the inclusion of a helical magnetic field complicates the character of the instability ({\em Nakamura et al.} 1993; {\em Fiege and Pudritz,} 2000). 
In the presence of a very strong toroidal field, for instance, the nature of the instability changes from gravitational to magnetic ({\em Fiege and Pudritz,} 2000). 

The evolution of massive, self-gravitating isothermal filaments was studied in detail by {\em Inutsuka and Miyama} (1992, 1997). 
They found that unless a large non-linear perturbation exists initially, a supercritical filament mainly collapses radially without fragmentation.  
Thus, the characteristic length scale, and hence the mass scale, is determined by the fragmentation of the filament at the bouncing epoch of the radial collapse. 
The bouncing of the radial collapse is expected when compressional heating 
starts to overwhelm radiative cooling, 
resulting in a significant increase in the gas temperature ($\gamma_{\rm eff} > 1$).  
The characteristic density, $\rho_{\rm crit}$, of this break-down of isothermality was studied with radiation hydrodynamics simulations 
by {\em Masunaga et al.} (1998) and its convenient expression for filamentary clouds was derived analytically by {\em Masunaga and Inutsuka} (1999).  
 
Once the radial collapse of a massive filament has been halted by an increase in temperature at higher density or some other mechanism, 
longitudinal motions of fragmentation modes are expected to become significant.  
These motions may create a number of dense cores located along the main axis of the filament, just as observed 
(e.g., Figs.~\ref{Aquila_coldens}a and ~\ref{growth_diag}b).  
Note that the characteristic core mass resulting from  fragmentation in this case is much smaller than the mass corresponding 
to the most unstable mode of the cylinder before the radial collapse. 
The resulting core masses are expected to depend on the initial amplitude distribution of density perturbations along the filament axis.  
Indeed, {\em Inutsuka} (2001) employed the so-called Press-Schechter formalism, a well-known method in cosmology ({\em Press and Schechter,} 1974), 
to predict the mass function of dense cores as a function of the initial power spectrum of fluctuations in mass per unit length.   
Figure~\ref{fig:PSMF} shows the typical time evolution of the core mass function. 
The evolution starts from initial perturbations of mass per unit length whose spectrum has a power-law index of $-1.5$, i.e., close to the Kolmogorov index of $-1.67$. 
This theoretical work suggests that if we know the power spectrum of fluctuations in line mass along 
filaments before gravitational fragmentation, 
we can predict the mass function of the resulting dense cores. 
Thus, it is of considerable interest to measure such power spectra for filaments that have not yet produced many cores (see Sect.~6.3 below). 

\bigskip
\noindent
\textbf{ 5.4 Roles of feedback effects }
\bigskip

Filamentary organization of the cloud plays a significant role in how important 
feedback effects can be.

\begin{figure}[!ht]
\epsscale{0.9}
\plotone{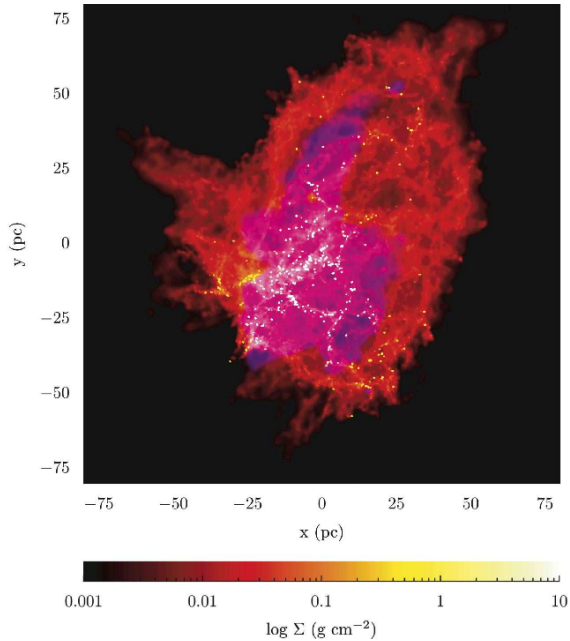}
\caption{\small 
Distribution of stars, cold gas filaments (in red), voids, and of the low-density ionized gas (in purple) that fills the cavities 
in the SPH numerical simulations of {\em Dale and Bonnell} (2011), in which a massive stellar cluster 
forms at the junction of a converging network of filaments, from the gravitational collapse of a turbulent molecular cloud.
  }
\label{Simulations2}  
\end{figure}

To understand the effect of radiative feedback from massive stars,  
consider the condition for having insufficient
ionizing radiative flux from some mass M undergoing an accretion flow to prevent the stalling or collapse of an HII region.
For spherical symmetry, 
{\em Walmsley} (1995) derived a simple formula for the critical accretion rate $\dot M_{crit}$ onto a source of mass $M$ 
and luminosity $L$ above which the ionizing photons from the source would be absorbed by the accreting
gas and cause the HII region to contract:
$ \dot M_{crit} > (4 \pi L G M m_H^2 / \alpha_B)^{1/2} $ where $\alpha_B$ is the 
hydrogenic recombination ($n>1$) coefficient.    Filamentary accretion offers a 
way out of this dilemma since it is intrinsically asymmetric.  
Ionizing flux can still escape the forming cluster in directions not dominated by 
the filamentary inflow ({\em Dale and Bonnell,} 2011).   Simulations show that the ionizing flux from forming massive clusters
escape into the low density cavities created by infall onto the filaments ({\em Dale and Bonnell,} 2011).  Such  simulations
have the character of a network of filaments and a ``froth'' of voids.  The filamentary 
flows themselves are far too strong to be disrupted by the radiation field.
{\em Dale and Bonnell} (2011 -- see their Fig.~12) find that the filamentary distribution of the dense star forming gas (about 3\% of the mass) 
is not disrupted in their simulation of a $10^6\, M_\odot$ clump seeded with supersonic Kolmogorov turbulence.  
Figure~\ref{Simulations2} shows the distribution of stars, filaments, voids, and low density ionized gas that fills the cavities 
in these simulations. 
The authors conclude 
that photoionization has little effect in disrupting the strong filamentary accretion 
flows that are building the clusters, 
although this seems to contradict 
the results of other work 
(e.g., {\em V\'azquez-Semadeni et. al.,} 2010; {\em Col\'\i n et. al.,} 2013). 
More work is needed to clarify the situation.

The feedback from outflows in a filament-dominated cloud can also have significant effects on stellar masses.   
Simulations by {\em Li et al.} (2010) show 
that outflows typically reduce the mass of stars by significant factors (e.g., from 1.3 to 0.5  $M_\odot $).   
Going further, {\em Wang et al.} (2010) assumed a simplified model for outflows in which the momentum injected from the sink particle into the 
surrounding gas $\Delta P = P_* \Delta M$  every time the sink particle has gained a mass of $\Delta M$.  The energy and momentum
are injected  in two highly collimated, bipolar jets.    
The outflows
from the cluster region are collectively powerful enough in these simulations to break up the very filaments that are strongly feeding material
into the cluster.  This severing of the filamentary accretion flow slows up the rate for cluster formation. In massive enough regions, however, 
the trickle of material that manages to avoid being impacted by these outflows can still feed the formation of massive stars.  

{\em Fall et al.} (2010 - see their Fig.~2) 
show that outflows play an important role for lower mass clouds and clusters ($ M < 10^4 M_\odot $).  
For higher masses and column densities $ \Sigma > 0.1 $ g cm$^{-2}$, however, 
radiation pressure seems to dominate over the feedback effects.  It will be interesting to see 
if this scaling still holds for filament-dominated clouds.  

\bigskip
\centerline{\textbf{ 6. CONFRONTING OBSERVATIONS AND THEORY}}
\medskip

\noindent
\textbf{ 6.1 A column density threshold for prestellar cores }
\medskip

\begin{figure*}
 \epsscale{2}
 \plotone{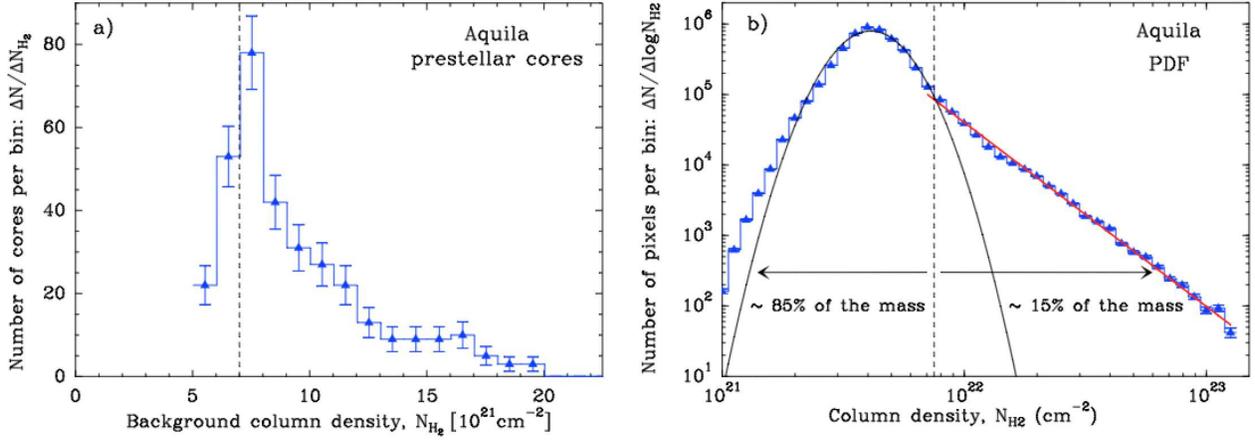}
 \vspace{-0.25cm}
 \caption{\small 
 {\bf (a)} Distribution of background column densities  for the 
candidate prestellar cores identified with $Herschel$ in the Aquila Rift complex (cf. {\em K\"onyves et al.,} 2010).
The vertical dashed line marks the column density or extinction threshold 
at $N_{\rm H_2}^{\rm back} \sim 7 \times 10^{21}$~cm$^{-2}$ or $A_V^{\rm back} \sim 8$ 
(also corresponding to $\Sigma_{\rm gas}^{\rm back} \sim $~130~$M_\odot \, {\rm pc}^{-2} $). 
{\bf (b)} Probability density function of column density in the Aquila cloud complex, based on the column density image
derived from $Herschel$ data ({\em Andr\'e et al.,} 2011; {\em Schneider et al.,} 2013). A log-normal fit at low column densities and a power-law fit at high 
column densities are superimposed. The vertical dashed line marks the same column density threshold as in the left panel.
  }
\label{threshold}       
 \end{figure*}

As already mentioned (e.g.,\S ~3.2), 
prestellar cores identified with $Herschel$ are preferentially found within the {\it densest filaments}, i.e., 
those with column densities exceeding $\sim 7 \times 10^{21}$~cm$^{-2}$ 
({\em Andr\'e et al.,} 2010 and Fig.~\ref{Aquila_coldens}a). 
In the Aquila region, for instance, the distribution of background cloud column densities for the population of prestellar cores 
shows a steep rise above $N_{\rm H_2}^{\rm back} \sim 5 \times 10^{21}$~cm$^{-2}$ (cf. Fig.~\ref{threshold}a) and is such that 
$\sim 90\% $ of the candidate bound cores are found above a background column density 
$N_{\rm H_2}^{\rm back} \sim 7 \times 10^{21}$~cm$^{-2}$, corresponding to a background 
visual extinction $A_V^{\rm back} \sim 8$. 
The $Herschel$ observations of 
Aquila 
therefore strongly support the existence of a column density or visual extinction 
threshold for the formation of prestellar cores at $A_V^{\rm back} \sim $~5--10, which had been suggested earlier based on 
ground-based studies of, e.g., the Taurus and Ophiuchus clouds (cf. {\em Onishi et al.,} 1998; {\em Johnstone et al.,} 2004). 
Interestingly, a very similar extinction threshold at $A_V^{\rm back} \sim 8$ is independently observed in the spatial 
distribution of YSOs with $Spitzer$ ({\em Heiderman et al.,} 2010; {\em Lada et al.,} 2010).  
In the Polaris flare cirrus, the $Herschel$ observations are also consistent with such an extinction threshold since the observed background column densities 
are all below $A_V^{\rm back} \sim 8$ and there are 
no or 
at most a handful of examples of bound prestellar cores in this cloud (e.g., {\em Ward-Thompson et al.,} 2010).
More generally, the results obtained with $Herschel$ in nearby clouds suggest that 
a fraction $f_{\rm pre} \sim 15\%$ of the total gas mass 
above the column density threshold is in the form of prestellar cores.

\bigskip
\noindent
\textbf{ 6.2  Theoretical interpretation of the threshold}
\bigskip

These $Herschel$ findings connect very well with theoretical expectations for the gravitational instability of filaments (cf. \S~5.1) 
and point to an {\it explanation} of the star formation threshold in terms of the filamentary structure of molecular clouds. 
Given the typical width $W_{\rm fil} \sim 0.1$~pc measured for 
filaments ({\em Arzoumanian et al.,} 2011; see Fig.~\ref{histo_width}) 
and the relation $M_{\rm line} \approx \Sigma_0 \times W_{\rm fil}$ between the central 
gas surface density $\Sigma_0$ 
and the mass per unit length $M_{\rm line}$ of a filament (cf. Appendix~A of {\em Andr\'e et al.,} 2010), 
the threshold at $A_V^{\rm back} \sim 8$ or $\Sigma_{\rm gas}^{\rm back} \sim $~130~$M_\odot \, {\rm pc}^{-2} $ 
corresponds to within a factor of $< 2$ to the critical mass per unit length $M_{\rm line, crit} = 2\, c_s^2/G \sim 16\, M_\odot \, {\rm pc}^{-1} $  
of nearly isothermal, long cylinders (see \S ~5.1 and {\em Inutsuka and Miyama,} 1997)  
for a 
typical 
gas temperature $T \sim 10$~K.
Thus, the core formation threshold approximately corresponds to the {\it threshold above which 
interstellar 
filaments are gravitationally unstable} 
({\em Andr\'e et al.,} 2010). 
Prestellar cores tend to be observed only above this threshold (cf. Fig.~\ref{Aquila_coldens}a \& 
Fig.~\ref{threshold}a) 
because they form out of a filamentary background and only 
filaments with $ M_{\rm line} > M_{\rm line, crit} $ are able to fragment into self-gravitating cores.

The column density threshold for core and star formation within filaments is not a sharp boundary but a smooth transition for several reasons. 
First, observations only provide information on the {\it projected} column density 
$\Sigma_{\rm obs} = \frac{1}{{\rm cos} \,i} \, \Sigma_{\rm int}$ of any given filament, where $i$ is the inclination angle to the plane of sky 
and $\Sigma_{\rm int}$ is the intrinsic column density of the filament (measured perpendicular to the long axis).
For a population of randomly oriented filaments with respect to the plane of sky, 
the net effect is that $\Sigma_{\rm obs} $ {\it overestimates} $\Sigma_{\rm int} $ by a factor  
$<\frac{1}{{\rm cos}\,i}>\, = \frac{\pi}{2}\,\sim1.57$ on average.  
Although systematic, this projection effect 
remains small and has little impact on the global classification of observed 
filamentary structures as supercritical or subcritical. 

Second, there is a (small) spread in the distribution of filament inner widths of about a factor of 2 on either side of 0.1~pc 
({\em Arzoumanian et al.,} 2011 -- cf. Fig.~\ref{histo_width}), implying a similar spread in the intrinsic column densities corresponding to the 
critical filament mass per unit length $M_{\rm line, crit} $. 

Third, interstellar filaments are not all exactly at $T = 10$~K and their internal velocity dispersion sometimes 
includes a small nonthermal component, $\sigma_{\rm NT}$, which must be accounted for in the evaluation 
of the critical or virial mass per unit length 
$M_{\rm line, vir} = 2\, \sigma_{\rm tot}^2/G $ 
(see \S ~5.1 and {\em Fiege and Pudritz,} 2000). 
The velocity dispersion measurements of {\em Arzoumanian et al.} (2013 -- cf. Fig.~\ref{velo_disp}b) confirm  
that there is a critical threshold 
in mass per unit length above which 
interstellar 
filaments are self-gravitating and below which they are unbound, and that the position of this 
threshold lies around $\sim \, $16--32$\, M_\odot \, {\rm pc}^{-1} $,
i.e., within a factor of 2 of the thermal value of the 
critical mass per unit length $M_{\rm line,crit}$ for $T = 10$~K. 
The results shown in Fig.~\ref{velo_disp}b emphasize the role played by the thermal critical mass per unit length $M_{\rm line,crit}$ 
in the evolution of  
filaments. Combined with the $Herschel$ findings summarized above 
and  illustrated in Figs.~\ref{Aquila_coldens}a 
and  \ref{threshold}a, they support the view that the gravitational fragmentation of filaments 
may control the bulk of core 
formation, 
at least in nearby Galactic clouds.

\bigskip
\noindent
\textbf{ 6.3  Filament fragmentation and the CMF/IMF peak }
\bigskip

Since most stars appear to form in filaments, the fragmentation of filaments at the threshold of gravitational instability is a plausible 
mechanism for the origin of (part of) the stellar IMF. We may expect local collapse into spheroidal protostellar cores to be controlled by the 
classical Jeans criterion $M \geq M_{\rm BE}$ where the Jeans or critical Bonnor-Ebert mass $M_{\rm BE}$ (e.g., {\em Bonnor,} 1956)  
is given by: 
\begin{equation}
  M_{\rm BE} \sim 1.3\, c_s^4 /G^2 \Sigma_{\rm cl}
\end{equation}
\noindent
or
$$ M_{\rm BE}  \sim 0.5\, M_\odot\, \times \left(T_{\rm eff} /10\, {\rm K}\right)^2 \times  \left(\Sigma_{\rm cl}/160\, M_\odot\, {\rm pc}^{-2}\right)^{-1}.\\ $$

\noindent
If we consider a quasi-equilibrium isothermal cylindrical filament on the verge of {\it global} radial collapse, 
it has a mass per unit length equal to the critical value $M_{\rm line, crit} = 2\, c_s^2/G$ ($\sim 16\, M_\odot \, {\rm pc}^{-1} $ for $T_{\rm eff} \sim 10$~K) and an effective 
diameter $D_{\rm flat, crit} = 2\, c_s^2/G\Sigma_0$ ($\sim 0.1\, $pc for $T_{\rm eff} \sim 10$~K and $\Sigma_0 \sim 160\, M_\odot\, {\rm pc}^{-2}$).
A segment of such a cylinder of length equal to $D_{\rm flat, crit}$ contains a mass 
$M_{\rm line, crit} \times D_{\rm flat, crit} = 4\, c_s^4/G^2\Sigma_0 \sim 3 \times M_{\rm BE} $ 
($\sim 1.6\, M_\odot$ for $T_{\rm eff} \sim 10$~K and $\Sigma_0 \sim 160\, M_\odot\, {\rm pc}^{-2}$) and is thus {\it locally} Jeans unstable. 
Since local collapse tends to be favored over global collapse in the case of filaments (e.g., {\em Pon et al.,} 2011 -- see \S~5.1), gravitational fragmentation 
into spheroidal cores is expected to occur along 
supercritical filaments, as indeed found in both numerical simulations (e.g., {\em Bastien et al.,} 1991; {\em Inutsuka and Miyama,} 1997) 
and $Herschel$ observations (see Fig.~\ref{Aquila_coldens}a and \S~6.1). 
Remarkably, 
the peak of the prestellar CMF at $ \sim 0.6\, M_\odot $ as observed in the Aquila cloud complex (cf. Fig.~\ref{Aquila_cmf}b) 
corresponds very well to the Bonnor-Ebert mass $ M_{\rm BE} \sim 0.5\, M_\odot $ 
within marginally critical filaments with $ M_{\rm line} \approx M_{\rm line, crit} \sim 16\, M_\odot \, {\rm pc}^{-1} $ and surface densities 
$\Sigma \approx \Sigma_{\rm gas}^{\rm crit} \sim 160\, M_\odot \, {\rm pc}^{-2} $. 
Likewise,  the median projected spacing $\sim 0.08$~pc observed between the prestellar cores of Aquila (cf. \S ~3.2) roughly matches the 
thermal Jeans length within marginally critical filaments.
All of this is consistent with the idea that gravitational fragmentation is the dominant physical mechanism generating prestellar cores 
within interstellar filaments. Furthermore, a typical prestellar core mass of $\sim 0.6\, M_\odot$  translates into a characteristic star 
or stellar system mass of $\sim 0.2\, M_\odot$, assuming a typical efficiency $ \epsilon_{\rm core} \sim 30\%$ (cf. \S ~3.4). 

\begin{figure}[ht]
\epsscale{1.0}
\vspace{-0.5cm}
\plotone{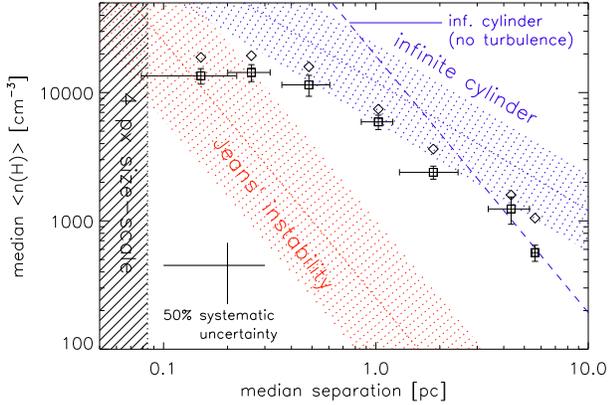}
\vspace{-0.75cm}
\caption{\small 
Median density of significant sub-structures as a function of separation in the high (2\arcsec  or 0.035~pc) resolution column 
density map of the massive, filamentary IRDC G11.11-0.12 by {\em Kainulainen et al.} (2013).
The red dotted line shows the relation expected from standard Jeans' instability arguments. 
The blue dotted line shows the relation predicted by the linear stability analysis of self-gravitating isothermal cylinders.
Note how the structure of G11 is consistent with 
global cylinder fragmentation on scales $\simgt 0.5$~pc and approaches local Jeans' fragmentation 
on scales $\simgt 0.2$~pc. 
(From {\em Kainulainen et al.,} 2013.)
  }
\label{spacings}  
\end{figure}

A subtle point needs to be made about the median spacing and mass of prestellar cores, however.
Namely, the values observed with $Herschel$ are a factor of $\simgt 4$ smaller than the characteristic 
core spacing and core mass predicted by linear stability analysis for infinitely long, $\sim 0.1$~pc-wide self-gravitating 
isothermal cylinders (see \S ~5.3). In their high-resolution infrared extinction study of the filamentary IRDC G11, 
{\em Kainulainen et al.} (2013) recently noted a similar mismatch between the typical spacing of small-scale ($\simlt 0.1$~pc) cores 
and the predictions of cylinder fragmentation theory. 
They were able to show, however, that the typical spacing of the larger-scale ($\simgt 0.5$~pc), 
lower-density sub-structures identified along the G11 filament was in reasonably good agreement with the global fragmentation properties 
expected for a self-gravitating cylinder (see Fig.~\ref{spacings}). This work suggests that on large scales the fragmentation properties 
of a self-gravitating filament are dominated by the most unstable mode of the global structure, while on small scales the fragmentation properties
depend primarily on local conditions and the local Jeans/Bonnor-Ebert criterion applies.

In any event, the $Herschel$ results tend to 
support the {\em Larson} (1985) interpretation of the peak of the IMF in terms 
of the typical Jeans mass in star-forming clouds. 
Overall, the $Herschel$ findings 
suggest that the gravitational fragmentation of supercritical filaments produces the peak of the prestellar CMF which, 
in turn, may account for the log-normal ``base'' (cf. {\em Bastian et al.,} 2010) of the IMF. 

It remains to be seen whether 
the bottom end of the IMF and the Salpeter power-law slope at the high-mass end can 
be also explained by filament fragmentation. 
Naively,  gravitational fragmentation should produce a narrow prestellar CMF, sharply peaked at the median thermal
Jeans mass. Note, however, that a small ($\sim 25\% $) fraction of prestellar cores do not appear to form along filaments (\S ~3.2). 
Furthermore,  a Salpeter power-law tail at high masses may result from filament 
fragmentation if turbulence has generated an appropriate field of initial density fluctuations within the filaments in the first place (cf. {\em Inutsuka,} 2001). 

Addressing the high-mass power-law tail of the IMF, 
{\em Inutsuka} (2001) has shown that if the power spectrum of initial density fluctuations along the filaments approaches 
$P(k) \equiv |\delta_k|^2 \propto k^{-1.5} $ then the CMF produced by gravitational fragmentation evolves toward 
$dN/dM \propto M^{-2.5}$ (see Fig.~\ref{fig:PSMF}),  similar to the Salpeter IMF  ($dN/dM_\star  \propto M_\star^{-2.35} $).
Interestingly, the power spectrum of column density fluctuations along the filaments observed with $Herschel$ in nearby clouds 
is typically $P(k) \propto k^{-1.6} $, which is close to the required spectrum ({\em Roy et al.,} in prep.). 

Alternatively, a CMF with a Salpeter power-law tail may result from the gravitational fragmentation of a population of filaments
with a distribution of supercritical masses per unit length. 
Observationally, the supercritical filaments observed as part of the $Herschel$ Gould Belt survey do seem to have a power-law
distribution of masses per unit length $dN/dM_{\rm line}  \propto M_{\rm line}^{-2.2}$ above $\sim 20\, M_\odot \, {\rm pc}^{-1} $ ({\em Arzoumanian et al.,} in prep.).
Since the width of the filaments is roughly constant ($W_{\rm fil}  \sim 0.1$~pc), the mass per unit length is directly proportional to 
the central surface density, $M_{\rm line} \sim \Sigma \times W_{\rm fil} $. 
Furthermore, the total velocity dispersion of these filaments increases roughly as $ \sigma_{\rm tot} \propto \Sigma^{0.5}$ ({\em Arzoumanian et al.,} 2013 -- see Fig.~\ref{velo_disp}b), 
which means that their effective temperature (including thermal and non-thermal motions) scales roughly as $T_{\rm eff} \propto \Sigma $. 
Hence $M_{\rm BE} \propto  \Sigma \propto M_{\rm line} $, and the observed distribution of masses per unit length directly translates into 
a power-law distribution of Bonnor-Ebert masses $dN/dM_{\rm BE}  \propto M_{\rm BE}^{-2.2}$ along supercritical filaments, which is also reminiscent 
of the Salpeter IMF.
\bigskip

\noindent
\textbf{ 6.4  A universal star formation law above the threshold? }
\smallskip

The realization that, at least in nearby clouds,  
prestellar core formation occurs primarily along gravitationally unstable filaments of roughly constant width $W_{\rm fil} \sim 0.1$~pc 
may also have implications for our understanding of star formation on global Galactic and extragalactic scales. 
Remarkably, the critical mass per unit length of a filament, $M_{\rm line, crit} = 2\, c_s^2/G$, depends only on gas temperature 
(i.e., $T \sim 10$~K for the bulk of molecular clouds, away from the immediate vicinity of massive stars) and is modified by 
only a factor of order unity for filaments with realistic levels of magnetization ({\em Fiege and Pudritz,} 2000 -- see Sect.~5.1). 
These simple conditions may set a quasi-universal threshold for star formation in the cold ISM of galaxies at $M_{\rm line, crit} \sim 16\, M_\odot \, {\rm pc}^{-1} $ in terms of 
filament mass per unit length, or $M_{\rm line, crit}/W_{\rm fil} \sim 160\, M_\odot \, {\rm pc}^{-2} $ in terms of gas surface density, 
or $M_{\rm line, crit}/W_{\rm fil}^2 \sim 1600\, M_\odot \, {\rm pc}^{-3} $ in terms of gas density 
(i.e., a 
number density 
$n_{H_2} \sim 2 \times 10^4\, {\rm cm}^{-3} $). 

While further work is needed to confirm that the width of interstellar filaments remains close to $W_{\rm fil} \sim 0.1$~pc 
in massive star-forming clouds beyond the Gould Belt, we note here that recent detailed studies of the RCW~36 and DR~21 ridges in Vela-C ($d \sim 0.7$~kpc) and 
Cygnus~X  ($d \sim 1.4$~kpc) are consistent with this hypothesis ({\em Hill et al.,} 2012; {\em Hennemann et al.,} 2012). 
As already pointed out in \S ~6.2, the threshold should be viewed as a smooth transition from non-star-forming to star-forming gas 
rather than as a sharp boundary. Furthermore, the above threshold corresponds to a {\it necessary} but not automatically {\it sufficient} condition 
for widespread star formation within filaments. In the extreme environmental conditions of the central molecular zone near the Galactic center, 
for instance, star formation appears to be largely suppressed above the threshold ({\em Longmore et al.,} 2013). 
In the bulk of the Galactic disk where more typical environmental conditions prevail, however, we may expect the above threshold in filament 
mass per unit length or (column) density to provide a fairly good selection of the gas directly participating in star formation. 

\begin{figure}[ht]
\setlength{\unitlength}{1mm}  
\noindent   
\begin{picture}(80,80) 
 \put(0,0){\includegraphics{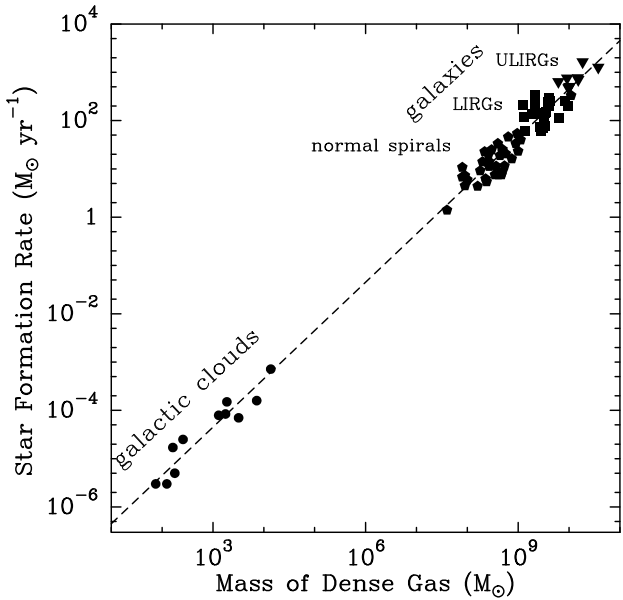}} 
\end{picture}
\vspace{-0.3cm} 
\caption{\small 
Relation between star formation rate (${\rm SFR}$) and mass of dense gas ($M_{\rm dense}$) for local molecular clouds ({\em Lada et al.,} 2010) 
and external galaxies ({\em Gao and Solomon,} 2004). $M_{\rm dense}$ is the mass of dense gas above the star formation threshold 
($n_{H_2} \sim 2 \times 10^4\, {\rm cm}^{-3} $ -- see text).
The dashed line going through the data points represents the linear relation 
${\rm SFR} = 4.5 \times 10^{-8}\,  M_\odot \, {\rm yr}^{-1}\, \times \left(M_{\rm dense}/M_\odot \right) $ inferred from the $Herschel$ 
results on prestellar cores in the Aquila cloud complex (see text and {\em K\"onyves et al.}, in prep.).
(Adapted from {\em Lada et al.,} 2012.)
  }
\label{sf_law}
\end{figure}

Recent near- and mid-infrared studies of the star formation rate as a function of gas surface density in both Galactic and extragalactic cloud complexes 
(e.g., {\em Heiderman et al.,} 2010; {\em Lada et al.,} 2010) show that the star formation rate tends to be linearly proportional to the 
mass of dense gas above a surface density threshold $\Sigma_{\rm gas}^{\rm th} \sim $~120--130~$M_\odot \, {\rm pc}^{-2} $ 
and drops to negligible values below $\Sigma_{\rm gas}^{\rm th} $ (see {\em Gao and Solomon,} 2004 for external galaxies). 
Note that this is essentially the {\it same} threshold as found with $Herschel$ for the formation of prestellar cores in nearby clouds (cf. \S ~6.1 and 
Figs.~\ref{threshold}a \& \ref{velo_disp}b).
Moreover, the relation between the star formation rate (${\rm SFR}$) and the mass of dense gas ($M_{\rm dense}$) above the threshold 
is estimated to be ${\rm SFR} = 4.6 \times 10^{-8}\,  M_\odot \, {\rm yr}^{-1}\, \times \left(M_{\rm dense}/M_\odot \right) $ in nearby clouds ({\em Lada et al.,} 2010), 
which is close to the relation ${\rm SFR} = 2 \times 10^{-8}\,  M_\odot \, {\rm yr}^{-1}\, \times \left(M_{\rm dense}/M_\odot \right) $ 
found by {\em Gao and Solomon} (2004) for galaxies (see Fig.~\ref{sf_law}).

Both of these values are very similar to the star formation rate per unit solar mass of dense gas 
${\rm SFR}/M_{\rm dense} = f_{\rm pre} \times \epsilon_{\rm core}/t_{\rm pre} \sim 0.15 \times 0.3/10^6 \sim 4.5 \times 10^{-8}\,  
{\rm yr}^{-1}$ 
that we may derive based on $Herschel$ in the Aquila cloud complex. 
For this estimate, we consider that only a fraction $f_{\rm pre} \sim 15\%$ of the gas mass above the column 
density threshold is in the form of prestellar cores (cf. \S ~6.1), that the local star formation efficiency at the level of an individual core is $ \epsilon_{\rm core} \sim 30\%$ (cf. \S ~3.4), 
and that the typical lifetime of the Aquila cores is $t_{\rm pre} \sim 10^6$~yr (cf. \S ~3.3 and {\em K\"onyves et al.}, in prep.). 
Despite relatively large uncertainties, the agreement with the extragalactic value of {\em Gao and Solomon} (2004) is surprisingly good, 
implying that there may well be a quasi-universal ``star formation law'' converting the gas of dense filaments 
into stars above the threshold (Fig.~\ref{sf_law} -- see also {\em Lada et al.,}  2012).
\bigskip

\noindent
\textbf{ 6.5  Origin of the characteristic inner width of filaments }
\smallskip

The fact that the same $\sim 0.1$~pc width is measured for low-density, subcritical filaments 
suggests that this characteristic scale is set by the physical process(es) producing the filamentary structure. 
Furthermore, at least in the case of  diffuse, gravitationally unbound clouds such as Polaris (Fig.~\ref{Polaris}), 
gravity is unlikely to be involved. 
As mentioned in Sect.~5.2, large-scale compression flows (turbulent or not) and the dissipation 
of the corresponding energy provide a potential mechanism. 
In the picture proposed by {\em Padoan et al.} (2001), for instance, the dissipation of turbulence occurs in shocks, 
and interstellar filaments correspond to dense, post-shock stagnation gas associated with compressed regions 
between interacting supersonic flows. One merit of this picture is that it can account qualitatively for 
the $\sim 0.1$~pc width.
The typical thickness of shock-compressed structures resulting from supersonic turbulence in the ISM is 
indeed expected to be 
roughly the sonic scale of the turbulence, i.e., $\sim 0.1$~pc in diffuse interstellar gas 
(cf. {\em Larson,} 1981; {\em Federrath et al.,} 2010, 
and discussion in {\em Arzoumanian et al.,} 2011). 
Direct evidence of the role of large-scale compressive flows has been found with $Herschel$ in the Pipe Nebula 
in the form of filaments 
with 
asymmetric column density profiles which most likely result from compression by the winds of the nearby 
Sco OB2 association ({\em Peretto et al.,} 2012). 

A more complete picture, 
proposed by {\em Hennebelle} (2013), is that 
filaments result from a combination of 
turbulent compression and shear 
(see {\em Hily-Blant \& Falgarone,} 2009 for an observed example of a diffuse CO filament in Polaris corresponding to 
a region of intense velocity shear). 
Interestingly, the filament width is comparable to the cutoff wavelength 
$\lambda_A \sim 0.1\, \rm{pc} \times (\frac{B}{10~\mu G}) \times (\frac{n_{H_2}}{10^3~\rm{cm}^{-3}})^{-1}$ 
below which MHD waves cannot propagate in the primarily neutral gas of molecular clouds (cf. {\em Mouschovias,} 1991), 
if the typical magnetic field strength is $B \sim 10\, \mu$G ({\em Crutcher,} 2012).
Hence the tentative suggestion that the filament width may be set by the dissipation mechanism of MHD waves 
due to ion-neutral friction ({\em Hennebelle,} 2013). 
Alternatively, the characteristic width may also be understood if interstellar filaments are formed as quasi-equilibrium structures 
in pressure balance with a typical ambient ISM pressure $P_{\rm ext}/k_{\rm B} {\sim} 2$$-$5$\times$$10^4 \, \rm{K\, cm}^{-3} $ 
({\em Fischera and Martin,} 2012; {\em Inutsuka et al.}, in prep.). 
Clearly, more work is needed to clarify the origin of the width of subcritical filaments.

That star-forming, supercritical filaments also maintain roughly constant inner widths $\sim 0.1$~pc
while evolving ({\em Arzoumanian et al.,} 2011 -- see Figs.~\ref{B211_prof} \& \ref{histo_width}) is even more surprising 
at first sight.
Indeed, supercritical filaments 
are unstable to radial collapse and are thus expected to undergo rapid radial contraction with time 
(see Sect.~5.1). 
The most likely solution to this paradox is that supercritical filaments are {\it accreting} additional background material while contracting. 
The increase in velocity dispersion with central column density observed for supercritical filaments 
({\em Arzoumanian et al.,} 2013 -- see Fig.~\ref{velo_disp}b)
is indeed suggestive of an increase in (virial) mass per unit length with time.   

As mentioned in Sect.~4.3, 
more direct observational evidence of this accretion process 
for supercritical filaments exists in 
several cases in the form of low-density striations or sub-filaments 
seen perpendicular to the main filaments 
and apparently feeding them from the side 
(see  Figs.~\ref{B211_fil_fibers},  ~\ref{B211_fil}, \&
~\ref{velo_disp}a). 
Accretion onto dense filaments is also seen in numerical simulations ({\em G\'omez and V\'azquez-Semadeni,} 2013). 
This process supplies gravitational energy to supercritical filaments which is then converted into turbulent kinetic energy 
(cf. {\em Heitsch et al.,} 2009; 
{\em Klessen and Hennebelle,} 2010) and may explain the observed increase in velocity dispersion with column density 
($ \sigma_{\rm tot} \propto {\Sigma_0}^{0.5}$ -- cf. Fig.~\ref{velo_disp}b). 
Indeed, the fine substructure and velocity-coherent ``fibers'' observed within $Herschel$ supercritical filaments 
(cf. Fig.~\ref{B211_fil_fibers} and {\em Hacar et al.,} 2013) may possibly be the manifestation of accretion-driven ``turbulence''. 
The central diameter of such accreting filaments is expected to be of order the effective Jeans length 
$D_{\rm J,eff} \sim 2\, \sigma_{\rm tot}^2/G\Sigma_0 $, which {\em Arzoumanian et al.} (2013) have shown to remain close to $\sim 0.1$~pc. 
Hence, through accretion of parent cloud material, supercritical filaments may keep roughly constant inner widths and remain 
in rough virial balance while contracting (see {\em Heitsch,} 2013a,b; {\em Hennebelle and Andr\'e,} 2013). 
This process may effectively prevent the global (radial) collapse of supercritical filaments 
and thus favor their fragmentation into cores (e.g., {\em Larson,} 2005), in agreement with the $Herschel$ results 
(see Figs.~\ref{Aquila_coldens}a \& \ref{growth_diag}b).

\bigskip
\centerline{\textbf{ 7. CONCLUSIONS: TOWARD A NEW }}
\centerline{\textbf{ PARADIGM FOR STAR FORMATION ?}}
\bigskip

The observational results 
summarized in \S ~2 to \S ~4  
provide key insight into the first phases of the star formation process. 
They emphasize the role of filaments and 
support a scenario 
in which the formation of prestellar cores 
occurs  in two main steps. 
First, the dissipation of kinetic energy in large-scale MHD flows (turbulent or not) 
appears to generate $\sim 0.1$ pc-wide filaments in the ISM.  
Second, the densest filaments  
fragment into prestellar cores by gravitational instability 
above the critical 
mass per unit length $M_{\rm line, crit}  \approx 16\, M_\odot \, {\rm pc}^{-1} $, equivalent  
to a critical (column) density threshold $ \Sigma_{\rm gas}^{\rm crit} \sim 160\, M_\odot \, {\rm pc}^{-2} $ ($A_V^{\rm crit} \sim 8$) 
or $ n_{\rm H_2}^{\rm crit} \sim 2 \times 10^4\,  {\rm cm}^{-3} $.

In contrast to the standard gravo-turbulent fragmentation picture (e.g., {\em MacLow and Klessen,} 2004), 
in which filaments are present but play no fundamental role, our proposed paradigm 
relies heavily on the unique properties of filamentary geometry, such as the existence of a critical 
mass per unit length for nearly isothermal filaments.

That the formation of filaments in the diffuse ISM represents the first step toward core and star formation is suggested by the filaments 
{\it already} being pervasive in a gravitationally unbound, non-star-forming cloud such as Polaris (cf. Fig.~\ref{Polaris}; {\em Hily-Blant and Falgarone,} 2007; 
{\em Men'shchikov et al.,} 2010; {\em Miville-Desch\^enes et al.,} 2010). Hence, many interstellar filaments are not produced by large-scale gravity and 
their formation must precede star formation. 

The second step appears to be the gravitational fragmentation of the densest filaments with supercritical masses 
per unit length ($ M_{\rm line} \ge M_{\rm line, crit} $) into 
prestellar cores (cf. \S ~6.2). 
In active star-forming regions such as the Aquila complex, most of the prestellar cores identified 
with $Herschel$ are indeed concentrated within supercritical filaments (cf. Fig.~\ref{Aquila_coldens}a). 
In contrast, in non-star-forming clouds such as Polaris, 
all of the filaments have subcritical masses per unit length and 
very few (if any) prestellar cores and no protostars 
are observed 
(cf. Fig.~\ref{Polaris}). 

The present scenario may explain the peak for the prestellar CMF 
and the base of the stellar IMF (see Sect.~6.3 and Fig.~\ref{Aquila_cmf}b). 
It partly accounts for the general inefficiency of the star formation process since, even in active star-forming complexes such as Aquila 
(Fig.~\ref{Aquila_coldens}), only a small fraction of the total gas mass ($\sim 15\%$ in the case of Aquila -- see Fig.~\ref{threshold}b) is 
above of the column density threshold, 
and only a small fraction $f_{\rm pre} \sim 15\%$ of the dense gas above the threshold is in the form of prestellar cores (see Sect.~6.1).
Therefore, the vast majority of the gas in a GMC ($\sim 98\% $ in the case of Aquila) does not participate in star formation at any given time 
(see also {\em Heiderman et al.,} 2010; {\em Evans,} 2011). 
Furthermore, the fact that essentially the same ``star formation law'' is observed above the column density threshold in both Galactic clouds 
and external galaxies (see Sect.~6.4; Fig.~\ref{sf_law}; {\em Lada et al.,} 2012) suggests the star formation scenario sketched above 
may well apply to the ISM of 
other galaxies. 

The results reviewed in this chapter 
are 
very encouraging as they 
tentatively 
point to a unified picture of star formation on GMC scales in both Galactic clouds and external galaxies. 
Much more work would be needed, however, to fully understand the origin of the characteristic width of interstellar filaments 
and to determine whether the same $\sim 0.1$ pc  width 
also 
holds beyond the 
clouds of the Gould Belt. 
In particular, confirming and refining the scenario proposed here will require follow-up observations to constrain the dynamics of the filaments 
imaged with $Herschel$ as well as detailed comparisons with numerical simulations of molecular cloud formation and evolution. 
ALMA and NOEMA 
will be instrumental in testing whether this scenario based on $Herschel$ 
results in 
nearby Galactic clouds is truly 
universal and applies to the ISM of all galaxies.

\textbf{ Acknowledgments.} 
We thank  
D. Arzoumanian, J. Kirk, V. K\"onyves, and P. Palmeirim for useful discussions 
and 
help with  
several figures. We also thank H. Beuther, E. Falgarone, and the referee for constructive comments. 
PhA is partially supported by the European Research Council under the European Union's Seventh 
Framework Programme (Grant Agreement no. 291294) and by the French National Research Agency (Grant no. ANR--11--BS56--0010).
REP is supported by a Discovery grant from the National Science and Engineering Research Council 
of Canada.

\bigskip

\centerline\textbf{ REFERENCES}
\bigskip
\parskip=0pt
{\small
\baselineskip=11pt
\refs Abergel, A. et al. 
1994, ApJ, 423, L59
\refs Alves, J. F., Lada, C. J., and Lada, E. A. 2001, Nature, 409, 159 
\refs Alves, J. F., Lombardi, M., \& Lada, C. J. 2007, A\&A, 462, L17
\refs Andr\'e, P., Belloche, A., Motte, F. et al. 
2007, A\&A, 472, 519
\refs Andr\'e, Ph.
et al. 2010, A\&A, 518, L102
\refs Andr\'e, Ph. et al. 
2011, in 
IAU Symp. 270, 
p. 255
\refs Andr{\' e}, P., Ward-Thompson, D., \& Barsony, M.\ 2000, in Protostars and Planets IV, Eds V. Mannings et al., p.59
\refs Arzoumanian, D.
et al. 2011, A\&A, 529, L6
\refs Arzoumanian, D. et al. 
2013, A\&A, 553, A119
\refs Ballesteros-Paredes, J. et al.
1999, ApJ, 527, 285
\refs Ballesteros-Paredes, J. et al. 
2003, ApJ, 592, 188
\refs Ballesteros-Paredes, J.
et al. 2006, ApJ, 637, 384
\refs Ballesteros-Paredes, J.
et al. 2011, MNRAS, 411, 65
\refs Bally, J., Langer, W.D., Stark, A.A. et al.
1987, ApJ, 312, L45
\refs Balsara, D. et al. 
2001, MNRAS, 327, 715
\refs Banerjee, R. et al. 
2006, MNRAS, 373, 1091
\refs Bastian, N., Covey, K.R., \& Meyer, M.R. 2010, ARA\&A, 48, 339
\refs Bastien, P., Arcoragi, J.-P., Benz, W. et al. 
1991, ApJ, 378, 255
\refs Basu, S., Ciolek, G. E., Dapp, W. B. et al.
2009, New A, 14, 483
\refs Bate, M. R., \& Bonnell, I. A. 2005, MNRAS, 356, 1201
\refs Bate, M. R. et al. 
2003, MNRAS, 339, 577
\refs Beichman, C. A.
et al. 1986, ApJ, 307, 337
\refs Belloche, A., Parise, B., Schuller, F. et al. 2011, A\&A, 535, A2
\refs Bergin, E.A., \& Tafalla, M. 2007, ARA\&A, 45, 339
\refs Beuther, H.
et al. 
2011, A\&A, 533, A17
\refs Bonnell, I. A. et al. 
2003, MNRAS, 343, 413
\refs Bonnor, W.B. 1956, MNRAS, 116, 351
\refs Bontemps, S., Andr\'e, P., K\"onyves, V. et al. 2010, A\&A, 518, L85
\refs Burkert, A., \& Hartmann, L. 2004, ApJ, 616, 288
\refs Cambr\'esy, L. 1999, A\&A, 345,  965
\refs Caselli, P., Benson, P. J., Myers, P. C. et al.
2002, ApJ, 572, 238
\refs Chabrier, G. 2005, ASSL, 327, 41
\refs Chapman, N.L.
et al. 2011, ApJ, 741, 21
\refs Chini, R.
et al. 1997, ApJ, 474, L135
\refs Col\'\i n, P. et al. 
2013, MNRAS, 435, 1701
\refs Commer\c con, B. et al.
2012, A\&A, 545, A98
\refs Contreras, Y., Rathborne, J., \& Garay, G.  2013, MNRAS, 433, 251
\refs Crutcher, R. M. 1999, ApJ, 520, 706
\refs Crutcher, R. M. 2012, ARA\&A, 50, 29
\refs Dale, J. E., \& Bonnell, I. A. 2011, MNRAS, 414, 321
\refs Di Francesco, J.
et al.
2007, in Protostars and Planets V, Eds. B. Reipurth et al. 
(Tucson: University of Arizona Press), p. 17
\refs Dunham, M.M.
et al.  2008, ApJS, 179, 249
\refs Egan, M. P.
et al. 1998, ApJ, 494, L199
\refs Elmegreen, B. G., \& Falgarone, E. 1996, ApJ, 471, 816
\refs Elmegreen, B. G., \& Scalo, J. 2004, ARA\&A, 42, 211
\refs Enoch, M. L., Young, K. E., Glenn, J. et al.  2008, ApJ, 684, 1240
\refs Evans, N.J.  1999, ARA\&A, 37, 311
\refs Evans, N.J.  2011, in 
IAU Symp. 270, 
p. 25
\refs Evans, N.J.
et al. 2009, ApJS, 181, 321
\refs Fall, S. M. et al. 
2010, ApJ, 710, L142
\refs Falgarone, E., Pety, J., \& Hily-Blant, P. 2009, A\&A, 507, 355
\refs Falgarone, E., Pety, J., \& Phillips, T. G. 2001, ApJ, 555, 178
\refs Federrath, C. et al. 
2010, A\&A, 512, A81
\refs Fiege, J.D., \& Pudritz, R.E. 2000,  MNRAS, 311, 85 
\refs Fischera, J., \& Martin, P.G. 2012, A\&A, 542, A77
\refs Gammie, C.F. et al. 
2003, ApJ, 592, 203
\refs Gao, Y., \& Solomon, P. 2004, ApJ, 606, 271
\refs Goldsmith, P.F.
et al. 2008, ApJ, 680, 428
\refs G\'omez, G. C., \& V\'azquez-Semadeni, E.  2013, 
astro-ph/1308.6298
\refs Gong, H., \& Ostriker, E.C. 2011, ApJ, 729, 120
\refs Goodman, A. A. et al. 
1990, ApJ, 359, 363
\refs Goodman, A. A. et al. 
1998, ApJ, 504, 223
\refs Goodwin, S. P., Nutter, D., Kroupa, P. et al. 
2008, A\&A, 477, 823
\refs Hacar, A., \& Tafalla, M. 2011, A\&A, 533, A34
\refs Hacar, A. et al. 
2013, A\&A, 554, A55
\refs Hartmann, L. 2002, ApJ, 578, 914
\refs Hatchell, J., \& Fuller, G. A. 2008, A\&A, 482, 855
\refs Hatchell, J. et al. 
2005, A\&A, 440, 151
\refs Heiderman, A.
et al. 2010, ApJ, 723, 1019
\refs Heitsch, F. 2013a, ApJ, 769, 115
\refs Heitsch, F. 2013b, ApJ, 776, 62
\refs Heitsch, F. et al. 
2009, ApJ, 704, 1735
\refs Hennebelle, P. 2013, A\&A, 556, A153
\refs Hennebelle, P., \& Andr\'e, Ph. 2013, A\&A, 560, A68
\refs Hennebelle, P., \& Chabrier, G. 2008, ApJ, 684, 395
\refs Hennemann, M.
et al. 2012, A\&A, 543, L3
\refs Henning, Th., Linz, H., Krause, O. et al. 2010, A\&A, 518, L95
\refs {Henshaw}, J.~D.
et al. 
2013, MNRAS, 428, 3425
\refs Hernandez, A. K., \& Tan, J. C. 2011, ApJ, 730, 44
\refs Heyer, M., Gong, H., Ostriker, E. \& Brunt, C. 2008, ApJ, 680, 420
\refs Heyer, M., Krawczyk, C., Duval, J. et al. 
2009, ApJ, 699, 1092
\refs Hill, T., Motte, F., Didelon, P. et al. 2011, A\&A, 533, A94
\refs Hill, T., Andr\'e, Ph., Arzoumanian, D. et al. 2012, A\&A, 548, L6
\refs Hily-Blant, P., \& Falgarone, E. 2007, A\&A, 469, 173
\refs Hily-Blant, P., \& Falgarone, E. 2009, A\&A, 500, L29
\refs Inoue, T., \& Inutsuka, S. 2008, ApJ, 687, 303
\refs Inoue, T., \& Inutsuka, S. 2009, ApJ, 704, 161
\refs Inoue, T., \& Inutsuka, S. 2012, ApJ, 759, 35
\refs Inutsuka, S. 2001, {ApJ}, 559, L149
\refs Inutsuka, S., \& Miyama, S.M. 1992, ApJ, 388, 392
\refs Inutsuka, S., \& Miyama, S.M. 1997, ApJ, 480, 681
\refs Jackson, J. M.
et al. 2010, ApJ, 719, L185
\refs Jessop, N. E., \& Ward-Thompson, D. 2000, MNRAS, 311, 63 
\refs Johnstone, D., \& Bally, J. 1999, ApJ, 510, L49
\refs Johnstone, D.
et al. 2000, ApJ, 545, 327
\refs Johnstone, D., Di Francesco, J., \& Kirk, H. 2004, ApJ, 611, L45
\refs Juvela, M., Ristorcelli, I., Pagani, L. et al. 2012, A\&A, 541, A12
\refs Kainulainen, J. et al. 
2013, A\&A, 557, A120
\refs Kawachi, T., \& Hanawa, T. 1998, PASJ, 50, 577
\refs Kevlahan, N., \& Pudritz, R. E. 2009, ApJ, 702, 39
\refs Kirk, H., Myers, P.C., Bourke, T.L. et al. 2013, ApJ, 766, 115
\refs Kirk, J.M. et al.
2005, MNRAS, 360, 1506
\refs Kirk, J.M.
et al. 2013, MNRAS, 432, 1424
\refs Klessen, R. S., \& Burkert, A. 2000, ApJS, 128, 287
\refs Klessen, R. S., \& Hennebelle, P. 2010, A\&A, 520, A17
\refs K\"onyves, V.
et al. 2010, A\&A, 518, L106
\refs Koyama, H. \& Inutsuka, S. 2000, ApJ, 532, 980
\refs Kramer, C., Stutzki, J., Rohrig, R. et al. 
1998, A\&A, 329, 249
\refs Kroupa, P. 2001, MNRAS, 322, 231
\refs Krumholz, M. R. et al. 
2007, ApJ, 656, 959
\refs Lada, C.J., Alves, J., \& Lada, E.A. 1999, ApJ, 512, 250
\refs Lada, C.J., Lombardi, M, \& Alves, J. 2010, {ApJ}, 724, 687
\refs Lada, C.J., Forbrich, J., Lombardi, M. et al. 
2012, {ApJ}, 745, 190
\refs Larson, R. B. 1969, MNRAS, 145, 271
\refs Larson, R.B., 1981, {MNRAS}, 194, 809
\refs Larson, R.B. 1985, MNRAS, 214, 379
\refs Larson, R. B. 2005, MNRAS, 359, 211
\refs Lee, C. W., \& Myers, P. C. 1999, ApJS, 123, 233
\refs Li, D., \& Goldsmith, P. F. 2012, ApJ, 756, 12
\refs Li, Z.-Y., Wang, P., Abel, T., \& Nakamura, F.  2010, ApJ, 720, L26
\refs Longmore, S. N.
et al. 2013, MNRAS, 429, 987
\refs MacLow, M.-M., \& Klessen, R.S. 2004, RvMP, 76, 125
\refs Malinen, J.
et al. 2012, A\&A, 544, A50
\refs Masunaga, H. et al. 
1998, ApJ, 495, 346
\refs Masunaga, H.,  \& Inutsuka, S. 1999, ApJ, 510, 822
\refs Matzner, C.D., \& McKee, C.F. 2000, ApJ, 545, 364
\refs Maury, A.
et al. 
2011, A\&A, 535, A77
\refs McClure-Griffiths, N. M.
et al. 2006, ApJ, 652, 1339
\refs McCrea, W.H. 1957, MNRAS, 117, 562
\refs Men'shchikov, A.
et al. 2010, A\&A, 518, L103
\refs Men'shchikov, A. et al. 
2012, A\&A, 542, A81
\refs Men'shchikov, A. 2013, A\&A, 560, A63
\refs Miettinen, O., \& Harju, J. 2010, A\&A, 520, A102
\refs Miville-Desch\^enes, M.-A.
et al. 2010, A\&A, 518, L104
\refs Miyama, S.M. et al. 
1987, Prog. Theor. Phys., 78, 1273
\refs Mizuno, A., Onishi, T., Yonekura, Y. et al. 1995, ApJ, 445, L161
\refs Molinari, S., Swinyard, B., Bally, J. et al. 2010, {A\&A}, 518, L100
\refs Molinari, S.
et al. 2011, A\&A, 530, A133
\refs Motte, F., Andr\'e, P., \& Neri, R. 1998, A\&A, 336, 150
\refs Motte, F. et al. 
2001, A\&A, 372, L41
\refs Motte, F., Bontemps, S., Schilke, P. et al. 2007, A\&A, 476, 1243
\refs Motte, F., Zavagno, A., Bontemps, S. et al. 2010, A\&A, 518, L77
\refs Mouschovias, T. Ch. 1991, ApJ, 373, 169
\refs Mouschovias, T. Ch., \& Paleologou, E. V. 1979, ApJ, 230, 204
\refs Myers, P. C. 1983, ApJ, 270, 105
\refs Myers, P.C.  2009, ApJ, 700, 1609
\refs Myers, P.C.  2011, ApJ, 735, 82
\refs Nagai, T., Inutsuka, S., \& Miyama, S. M. 1998, ApJ, 506, 306
\refs Nagasawa, M.  1987, Prog. Theor. Phys., 77, 635
\refs Nakano, T, \& Nakamura, T. 1978, PASJ, 30, 671
\refs Nakamura, F., \& Umemura,  M. 1999, ApJ, 515, 239
\refs Nakamura, F., Hanawa, T., \& Nakano, T. 1993, PASJ, 45, 551
\refs Nutter, D. et al. 
2008, MNRAS, 384, 755
\refs Onishi, T., Mizuno, A., Kawamura, A. et al. 
1998, ApJ, 502, 296
\refs {Ostriker}, J. 1964, ApJ, 140, 1056
\refs {Ostriker}, E. C. et al. 
1999, ApJ, 513, 259
\refs Padoan, P. \& Nordlund, A. 2002, ApJ, 576, 870
\refs Padoan, P., Juvela, M., Goodman, A. et al. 
2001, ApJ, 553, 227
\refs Pagani, L., Ristorcelli, I., Boudet, N. et al. 2010, A\&A, 512, A3
\refs Palmeirim, P., Andr\'e, Ph., Kirk, J. et al. 2013, A\&A, 550, A38
\refs Passot, T. et al. 
1995, ApJ, 455, 536
\refs P\'erault, M., Omont, A., Simon, G. et al. 1996, A\&A, 315, L165
\refs Peretto, N., \& Fuller, G.A. 2009, A\&A, 505, 405
\refs Peretto, N., Andr{\' e}, Ph., K\"onyves, V. et al.\ 2012, A\&A, 541, A63
\refs Peretto, N.
et al. 2013, A\&A, 555, A112
\refs Pezzuto, S., Elia, D., Schisano, E. et al.\ 2012, A\&A, 547, A54
\refs Pilbratt, G.L.
et al. 2010, A\&A, 518, L1
\refs Pineda, J., Goodman, A., Arce, H. et al. 2010, ApJ, 712, L116
\refs Pineda, J., Goodman, A., Arce, H. et al. 2011, ApJ, 739, L2
\refs {Pineda}, J., {Rosolowsky}, E., \& {Goodman}, A. 2009, ApJ, 699, L134
\refs Polychroni, D., Schisano, E., Elia, D. et al. 2013, ApJL, 777, L33
\refs Pon, A., Johnstone, D., \& Heitsch, F. 2011, ApJ, 740, 88
\refs Pon, A., Toal\'a, J. A., Johnstone, D. et al. 
2012, ApJ, 756, 145
\refs Porter, D. et al. 
1994, Phys. Fluids, 6, 2133
\refs Press, W. \& Schechter, P. 1974, ApJ, 187, 425
\refs Pudritz, R. E., \& Kevlahan, N. K.-R. 2013, Phil. Trans. R.~Soc.~A., 371, 20120248
\refs {Rosolowsky}, E.~W.
et al.  2008, ApJ, 679, 1338
\refs Reid, M. A., Wadsley, J., Petitclerc, N. et al. 
2010, ApJ, 719, 561
\refs Russeil, D.
et al. 2013, A\&A, 554, A42
\refs Saigo, K. \& Tomisaka, K. 2011, ApJ, 728, 78
\refs Sandell, G., \& Knee, L. B. G. 2001, ApJ, 546, L49
\refs Schnee, S., Caselli, P., Goodman, A. et al. 2007, ApJ, 671, 1839
\refs {Schneider}, N., {Andr\'e}, P., {K\"onyves}, V. et al. 2013, ApJL, 766, L17
\refs {Schneider}, N.
et al. 
2010, A\&A, 520, A49
\refs Schneider, N.
et al. 2012, A\&A, 540, L11
\refs Schneider, S. \& Elmegreen, B.G. 1979, ApJS, 41, 87
\refs Shu, F. 1977, ApJ, 214, 488
\refs Simpson, R. J. et al. 
2011, MNRAS, 417, 216
\refs Smith, R. J.
et al. 2011, MNRAS, 411, 1354 
\refs Smith, R. J., Shetty, R., Stutz, A. M. et al. 
2012, ApJ, 750, 64
\refs Sousbie, T.  2011, MNRAS,  414, 350
\refs Stanke, T. et al. 
2006, A\&A, 447, 609
\refs Starck, J. L., Donoho, D. L., Cand\`es, E. J.\ 2003, A\&A, 398, 785
\refs Stodolkiewicz, J. S. 1963, AcA, 13, 30
\refs Stutzki, J., \& G\"usten, R. 1990, ApJ, 356, 513
\refs Sugitani, K.
et al. 2011, ApJ, 734, 63
\refs Tackenberg, J.
et al. 2012, A\&A, 540, A113
\refs Tafalla, M. et al. 
2004, A\&A, 416, 191
\refs Tilley, D.A., \& Pudritz, R.E. 2004, MNRAS, 353, 769
\refs Tilley, D.A., \& Pudritz, R.E. 2007, MNRAS, 382, 73
\refs Toal\'a, J. A. et al. 
2012, ApJ, 744, 190
\refs Tohline, J.E. 1982, Fund. of Cos. Phys., 8, 1
\refs V\'azquez-Semadeni, E. 1994, ApJ, 423, 681
\refs V\'azquez-Semadeni, E.
et al. 
2007, ApJ, 657, 870
\refs V\'azquez-Semadeni, E.
et al. 2010, ApJ, 715, 1302
\refs Walmsley, M. 1995, RMxAC, 1, 137
\refs Wang, P., Li, Z.-Y., Abel, T., \& Nakamura, F. 2010, ApJ, 709, 27
\refs Ward-Thompson, D. et al. 
1994, MNRAS, 268, 276
\refs Ward-Thompson, D. et al. 
2007, in Protostars and Planets V, Eds. B. Reipurth et al. 
(Tucson: University of Arizona Press), p. 33
\refs Ward-Thompson, D.
et al. 2010, A\&A, 518, L92
\refs Whitworth, A.P. et al. 
1996, MNRAS, 283, 1061
\refs Whitworth, A.P., \& Ward-Thompson, D. 2001, ApJ, 547, 317
\refs Williams, J.P., de Geus, E.J., \& Blitz, L. 1994, ApJ, 428, 693	

\end{document}